%% file: HancovaGDD2021v2.tex
\documentclass[]{interact}

\usepackage{epstopdf}
\usepackage[caption=false]{subfig}

\usepackage{bigints}
\usepackage{url}
\usepackage{mathtools}

\usepackage[numbers,sort&compress]{natbib}
\bibpunct[, ]{[}{]}{,}{n}{,}{,}
\renewcommand\bibfont{\fontsize{10}{12}\selectfont}
\makeatletter
\def\NAT@def@citea{\def\@citea{\NAT@separator}}
\makeatother

\theoremstyle{plain}
\newtheorem{theorem}{Theorem}[section]

\theoremstyle{definition}
\newtheorem{definition}[theorem]{Definition}

\theoremstyle{remark}
\newtheorem{remark}{Remark}

\input{OwnSymbolsCommands}

\begin{document}

\articletype{original article}

\title{A practical, effective calculation of gamma difference distributions with open data science tools}

\author{
	\name{Martina Han\v{c}ov\'a\textsuperscript{a}, Andrej Gajdo\v{s}\textsuperscript{a} \thanks{CONTACT Martina Han\v{c}ov\'a. Email: martina.hancova@upjs.sk} and Jozef Han\v{c}\textsuperscript{b}}
	\affil{\textsuperscript{a}Institute of Mathematics, Faculty of Science, Pavol Jozef \v{S}af\'arik University in Ko\v{s}ice, Slovakia; \textsuperscript{b}Institute of Physics, Faculty of Science, Pavol Jozef \v{S}af\'arik University in Ko\v{s}ice, Slovakia }
}

\maketitle

\begin{abstract}

At present, there is still no officially accepted and extensively verified implementation of computing the gamma difference distribution allowing unequal shape parameters. We explore four computational ways of the gamma difference distribution with the different shape parameters resulting from time series kriging, a forecasting approach based on the best linear unbiased prediction, and linear mixed models.
The results of our numerical study, with emphasis on using open data science tools, demonstrate that our open tool implemented in high-performance Python(with Numba) is exponentially fast, highly accurate, and very reliable. It combines numerical inversion of the characteristic function and 
the trapezoidal rule with the double exponential oscillatory transformation (DE quadrature). At the double 53-bit precision, our tool outperformed the speed of the analytical computation based on Tricomi's $U(a, b, z)$ function in CAS software (commercial Mathematica, open SageMath) by 1.5-2 orders. At the default precision of scientific numerical computational tools, it exceeded open SciPy, NumPy, and commercial MATLAB 5-10 times. The potential future application of our tool for a mixture of characteristic functions could open new possibilities for fast data analysis based on exact probability distributions in areas like multidimensional statistics, measurement uncertainty analysis in metrology as well as in financial mathematics and risk analysis.

\end{abstract}

\begin{keywords}
	 numerical inversion of the characteristic function; double exponential quadrature; computational tools; high-performance Python; econometrics; time series; kriging
\end{keywords}

\section{Introduction}
\leavevmode

The distribution of the difference of two independent gamma random variables with the same shape parameters belonging to the Laplace distribution family provides stochastic models with extensive applications in various areas such as economics, finance, communications, engineering, biology, physics, or geosciences \cite{kotz_laplace_2001,kozubowski_laplace_2012}. 

As Klar \cite{klar_note_2015} pointed out, such gamma difference, but with unequal shape parameters, has received much less attention, and before 2015 it was mentioned very rarely in the research or literature. Over the last five years, thanks to Klar's review article, the first applications have begun to appear, and research has become more intensive. \\

Now we can find applications of Klar's results on the gamma difference distribution (with different shape parameters), or shortly $\GDD$, in controlling a measurement accuracy of optical detectors \cite{hendrickson_centralized_2017}, detecting radar sensor threshold \cite{ranney_efficient_2021}, setting optimal performance of wifi networks \cite{khan_wirelessly_2017}, chemotherapy cancer treatment \cite{sekhavati_dynamic_2015}, detecting eye glaucoma \cite{belghith_learning_2015}, stochastic modeling of the forest composed by point-to-line geodesics \cite{lopez_geodesic_2017}, or using lidar time series in forest mapping \cite{dial_estimating_2021}.

Regarding computing $\GDD$, we can notice that in Klar \cite{klar_note_2015} there were only several vague recommendations\footnote{In the paper there is no reference what computational software was used for $\GDD$ plots or computations. According to the visual appearance of the graphs we assume that the author used MATLAB.} on how to compute $\mathcal{GDD}$ efficiently and practically. Up to date, we do not know any officially accepted and extensively verified implementation for $\GDD$ allowing unequal shape parameters. The choice of an appropriate digital tool for scientific computing is still silently left to a potential $\GDD$ user. However, the need for accurate, reliable, numerically stable, and fast computations arises naturally in any real data analysis or computational research (e.g. Monte Carlo or bootstrap methods). 

Therefore, in this work we investigate the four principal computational ways for $ \GDD$ in the plethora of currently available computing tools. We will pay special attention to open digital tools based on programming languages Python and R, which became significant during the last decade, in the light of enormous advances of open data science. 
It is worth to mention that open data science provides  open digital tools available to everybody in the statistical community, allowing easy and very efficient reproducibility, collaboration, and communication \cite{lowndes_our_2017,chambers_breaking_2017}.

The paper contains the next two main sections. In sec. 2 we outline the theoretical background for computing the probability density function (\textit{pdf}) or the cumulative distribution function (\textit{cdf}) of the gamma difference distribution. Our recapitulation revisits and in some details expands the published theory, introduced and clarified in the mentioned work of Klar \cite{klar_note_2015}. 

The theoretical background also provides a necessary starting point for our numerical study presented in the following sec. 3. The section states an existing computational problem, a review of current computing tools used by us in exploring the effectiveness of analytic and numerical methods dealing with computing $\GDD$, including our own several proposals. 
The numerical study, based on real data, also demonstrates another application of $\GDD$ dealing with time series econometrics based on kriging methodology \cite{gajdos_kriging_2017, hancova_estimating_2020}. Finally, for the sake of paper readability, in appendix A we present a comprehensive list of all formulas used in theoretical considerations. In appendix B we report some extra mathematical and statistical details how $\GDD$ appears in our time series application. In appendix C we summarize all open digital tools we applied in our numerical study.  

\vspace{-6pt}
\Section{Gamma difference distribution and its computation}
\vspace{-6pt}
\subsection{Three basic ways for computing $\GDD$}

The aim of this section is to revisit, reformulate in more compact way and supplement theory and formulas in  Klar \cite{klar_note_2015}, which lead to three basic ways for computing $\GDD$
\begin{enumerate}
	\item a numerical quadrature of the pdf convolution integral,
	\item a closed-form analytic expression of the pdf convolution integral \\ via confluent hypergeometric functions,
	\item a numerical quadrature of the cdf integral.
\end{enumerate}

\noindent In the beginning, we adopt the following definition and notation of the univariate gamma difference distribution from Klar \cite{klar_note_2015}.

\begin{definition}[\textbf{\textit{Gamma difference distribution}}]
\leavevmode\\
\itshape
Let us assume that two independent random variables $X_1, X_2$ have a gamma distribution with corresponding parameters $\a_j >0,\,\b_j>0, j = 1,2$ 
$$
X_j\sim\Gm(\a_j,\b_j), \, j =1,2.
$$ 
Then we call the distribution of the difference $X \equiv X_1-X_2$ a \textit{gamma difference distribution} with parameters $\a_1,\b_1,\a_2,\b_2$ or $\GDD(\a_1,\b_1,\a_2,\b_2)$ for short. 
\end{definition}

\vspace{-12pt}

\subsection*{(1) The pdf convolution integral}

Combining the following parametrization of the pdf of gamma random variables $X_j, \, j=1,2$ with positive support \cite{mittelhammer_mathematical_2013}
$$f_j(x)=\frac{\b_j^{\a_j}}{\Gamma(\a_j)}x^{\a_j-1}e^{-\b_j x}, 
\qquad x\in(0,\infty),\,\a_j>0,\,\b_j>0$$
\big($\Gamma(\a_j)$ denotes the gamma function $\Gamma(\a_j) = \int_{0}^{\infty}e^{-x}x^{\a_j-1} dx, \,  j=1,2$\big) and the well-known convolution formula \cite{shorack_probability_2017} to gamma distributions $X_j$ in opposite directions, we obtain for the pdf $f(.)$ of $X=X_1-X_2$ the following expression \cite[eq. (4) in][]{klar_note_2015}
\begin{equation}\label{eq:GDDdensconv}
f(z)=
\dfrac{\b_1^{\a_1}\b_2^{\a_2}}{\Gamma(\a_1)\Gamma(\a_2)}
\begin{cases} 
e^{-\b_1z}\bigints_{-z}^{\infty} x^{\a_2-1}(x+z)^{\a_1-1}e^{-\b x} dx, & z<0, \\[12pt]
e^{\b_2z}\bigints_{z}^{\infty} x^{\a_1-1}(x-z)^{\a_2-1}e^{-\b x} dx, & z>0, 
\end{cases}
\end{equation} 
where $\b \equiv \b_1+\b_2.$

\vspace{-6pt}
\subsubsection*{Special case: $f(z), z=0$}
The value of $f$ at $z=0$, not specifically mentioned in Klar \cite{klar_note_2015}, requires a little bit more attention. In this case, employing basic properties of the gamma function $\Gamma(z)$ and standard integral convergence criteria \cite{mittelhammer_mathematical_2013}, we can quickly get  ($\a\equiv \a_1+\a_2$) 
$$\displaystyle f(0) = \lim_{z\to 0^{-}} f(z) = \lim_{z\to 0^{+}} f(z) =
\begin{cases}
\infty, & 0<\a\leq 1, \\[12pt]
\dfrac{\b_1^{\a_1}\b_2^{\a_2}}{\b^{\a-1}} \dfrac{\Gamma(\a-1)}{\Gamma(\a_1)\Gamma(\a_2)}, &  1<\a.
\end{cases}  $$
This result can  also be obtained as a consequence of \eqref{form:Uabzint} and \eqref{form:Uab0}. Mathematically equivalent result expressed
via the beta function can be found as Lemma 2.1 in \citep{hendrickson_inverse_2019}.
 
\vspace{-12pt}

\subsection*{(2) A closed-form of the pdf convolution integral} 

Integrals in \eqref{eq:GDDdensconv} can be expressed as closed forms in terms of special functions, particularly confluent hypergeometric functions. We rewrite Klar's form \cite[eq. (5), p.~4]{klar_note_2015} to three different closed forms, since not all types of hypergeometric functions are available in statistical or mathematical software. Another reason consists in the fact that the special functions can have different computer implementations with respect to speed or reliability of computations. Simultaneously we introduce a more compact and readable notation compared to Klar \cite{klar_note_2015}, inspired by Hendrickson \cite{hendrickson_centralized_2017,hendrickson_inverse_2019}.

\begin{remark}\label{rem:DLMF}\textit{Special functions}
\small
\leavevmode\\[6pt]	
In the area of special functions, we cite and rely on the unique online project \textit{The Digital Library of Mathematical Functions} (DLMF \cite{dlmf_nist_2021}), which completely revised, updated, and expanded one of the most important and cited mathematical handbooks --- Abramowitz and Stegun's Handbook from 1964 \cite{abramowitz_handbook_2014}. In some cases, our references are also supplemented by another classic handbook from Gradshteyn and Ryzhik \cite{gradshteyn_table_2007}. Based on these works, we have written used key formulas in appendix A for not difficult following of our theoretical arguments. We do not provide definitions of used special functions as they can be easily found in given references.
\end{remark}

\vspace{-12pt}
\subsubsection*{Whittaker's function $W_{\upkappa,\upmu}(z)$} 
If we apply the integral identity \eqref{form: WkzGDDconv} connecting the pdf convolution integral \eqref{eq:GDDdensconv} and Whittaker's confluent hypergeometric functions $W_{\upkappa,\upmu}(z)$, we get
\begin{equation}\label{eq:GDD_densW}
f(z)=
\dfrac{\b_1^{\a_1}\b_2^{\a_2}}{\b^{\a/2}}\begin{cases} 
\dfrac{(-z)^{\a/2-1}}{\Gamma(\a_2)}e^{(-z)(\b_1-\b_2)/2} \,
W_{\textstyle\frac{\a_2-\a_1}{2},\frac{1-\a}{2}}\big(\!-\!z\b\big), & z< 0, \\[12pt]
\begin{array}{cc}
\frac{\Gamma(\a-1)}{\Gamma(\a_1)\Gamma(\a_2)}\b^{1-\a/2}, & \qquad \scriptstyle 1 < \a,  \\
\infty, & \scriptstyle \qquad  0 < \a \leq 1,
\end{array} & {z = 0,} \\[12pt] 
\dfrac{z^{\a/2-1}}{\Gamma(\a_1)} e^{z(\b_2-\b_1)/2}  \,
W_{\textstyle\frac{\a_1-\a_2}{2},\frac{1-\a}{2}}\big(z\b\big), & z>0. 
\end{cases}
\end{equation}
where $\a = \a_1+\a_2$ and $\b = \b_1+\b_2$. Expression \eqref{eq:GDD_densW} represents an expanded but also the less complicated version of eq. (5) in Klar \cite[p.~4]{klar_note_2015}, originally first published and derived in Mathai \cite{mathai_noncentral_1993} as theorem 2.1.

\subsubsection*{Tricomi's function $U(a,b,z)$}
Using relation \eqref{form:Wkmz} between Whittaker's $W_{\upkappa,\upmu}(z)$  and Tricomi's (Kummer's) $U(a,b,z)$ confluent hypergeometric functions, we can write a new closed form of \eqref{eq:GDDdensconv}
\begin{equation}\label{eq:GDD_densU}
f(z)=
\dfrac{\b_1^{\a_1}\b_2^{\a_2}}{\b^{\a-1}}\begin{cases} 
\dfrac{e^{z\b_2}}{\Gamma(\a_2)} U(1-\a_2, 2- \a, -z\b), & z< 0, \\[12pt]\begin{array}{cc}
\frac{\Gamma(\a-1)}{\Gamma(\a_1)\Gamma(\a_2)}, & \qquad \scriptstyle 1 < \a,  \\
\infty, & \scriptstyle \qquad  0 < \a \leq 1,
\end{array} & {z = 0,} \\[12pt] 
\dfrac{ e^{-z\b_1}}{\Gamma(\a_1)} U(1-\a_1, 2-\a, z\b), & z>0, 
\end{cases}
\end{equation}

This important form of $f(x)$, which can also be found in \cite{hendrickson_centralized_2017}, is not only much more economic and comprehensible as \eqref{eq:GDD_densW} or mentioned, more ``messy'' eq. (5) in \cite{klar_note_2015}, but it contains Tricomi's function \eqref{form:Uabz2F0}, one of the most commonly used hypergeometric functions with a wide variety of applications\footnote{e.g. in finance (Asian options), in genetics (gene-frequency analysis) or physical sciences (wave equation, scattering), see a full list of references in \cite{pearson_numerical_2017}}. 
Such practical significance of Tricomi's function has attracted enough attention to provide its computation by many open and commercial digital tools.

On the other hand, any reliable, fast, and rigorous computational implementation  of $U(a,b,z)$ requires a carefully mastered roadmap of different numerical methods and techniques from Taylor or asymptotic series via numerical quadratures to recurrence relations \cite{pearson_numerical_2017,gil_numerical_2007}. In the case of some parameters $a, b$ even today's best and most powerful digital tools and software can fail \cite{johansson_computing_2019}. Since $W_{\upkappa,\upmu}(z)$ are usually derived from $U(a,b,z)$ by \eqref{form:Wkmz}, any computational failure is almost surely caused by a failure of $U(a,b,z)$.

\vspace{-6pt}
\subsubsection*{Generalized hypergeometric function ${}_{2}F_{0}(a,b;z)$}
Finally, using the well-known connection \eqref{form:Uabz2F0}  between $U(a,b,z)$ and generalized hypergeometric function  ${}_{2}F_{0}(a,b;z)$, we have the third closed version of $f(x)$ 
\begin{equation}\label{eq:GDD_dens2F0}
f(z)=
\dfrac{\b_1^{\a_1}\b_2^{\a_2}}{\b^{\a}}\begin{cases} 
\dfrac{(-z)^{\a_2-1}e^{z\b_2}}{\b^{-\a_2}\Gamma(\a_2)} \, {}_2F_0\left(\begin{matrix} {\a_1},1-{\a_2} \\  \end{matrix} ; \frac{1}{{\b} z}\right), & z< 0, \\[12pt]\begin{array}{cc}
\frac{\Gamma(\a-1)}{\Gamma(\a_1)\Gamma(\a_2)}\b, & \qquad \scriptstyle 1 < \a,  \\
\infty, & \scriptstyle \qquad  0 < \a \leq 1,
\end{array} & {z = 0,} \\[12pt] 
\dfrac{(-z)^{\a_1-1}e^{-z\b_1}}{\b^{-\a_1}\Gamma(\a_1)} \, {}_2F_0\left(\begin{matrix} {1-\a_1},{\a_2} \\  \end{matrix} ; \frac{1}{{\b} z}\right), & z>0, 
\end{cases}
\end{equation}
This form is computationally interesting, because  ${}_{2}F_{0}(a,b;z)$ belongs with ${}_{0}F_{1}, {}_{1}F_{1},  {}_{2}F_{1}$ to special cases of the generalized hypergeometric series \eqref{form:genhypseries}, mostly implemented in computer software. 
Here we remind that expressions  \eqref{form:Uabz2F0},  \eqref{form:Wkmz}, \eqref{eq:GDD_dens2F0} with ${}_{2}F_{0}$ are valid in the sense of the Borel integral summability 
(\cite[sec. 6.1]{johansson_computing_2019}, \cite{byattsmith_borel_1999}, \cite[sec. 4.13]{misra_introduction_2016}).

\vspace{-12pt}
\subsection*{(3) The cdf integral}

According to \cite{klar_note_2015}, it is straightforward to get the following integral expression for the $\GDD$ cdf $F(.)$  from the formula  $P(X_1-X_2\le t) = \int_0^\infty P(X_1 \le x+t)f_2(x)dx$ 
\begin{equation}\label{eq:GDD_DF} 
F(t)=\dfrac{\b_2^{\a_2}}{\Gamma(\a_1)\Gamma(\a_2)}\bigintsss_{\mathrm{max}\{0,-t\}}^{\infty}x^{\a_2-1}e^{-\b_2x}
\gamma(\a_1,\b_1(x+t))dx,\,\,\,\,\,t\in\Rv{},
\end{equation}
where $\gamma(a,x)\equiv\int_{0}^{x}t^{a-1}e^{-t}dt$ is the lower incomplete gamma function. The same result can be obtained \cite{hendrickson_centralized_2017} by integrating the 
joint density $f(x_1,x_2) = f_1(x_1) \cdot f_2(x_2)$ of $X$. Finally, any numerical value of $f(x)$ can be computed with the help of the standard numerical differentiation of $F(x)$ \cite[see e.g. the five point formula,][]{turner_applied_2018}.

\begin{remark}\label{rem:Hendrickson}\textit{The lower integration limit}
	\small
	\leavevmode\\[6pt]
	It is important to point out that the formula (19) in  \cite{hendrickson_centralized_2017} derived by Hendrickson has the incorrect lower bound $0$, instead of correct $\mathrm{max}\{0,-t\}$. For negative $t$, the formula would lead to meaningless results caused by complex values of the lower incomplete gamma function. 
\end{remark}

\vspace{-12pt}
\subsubsection*{Special case: $F(z), z=0$}
In practice, a $\GDD$ difference $X$ serves many times as a statistical model for a real quantity whose meaning imposes naturally some limiting constrain conditions on its values. The typical example of constraints as prior information is nonnegativity. 

In such case, the cdf integral \eqref{eq:GDD_DF} also provides us a useful exact analytic expression for the probability of any unacceptable negative result of $X$
\begin{equation}\label{eq:F0}
\begin{gathered}
F(0) = \dfrac{\b_1^{\a_1}\b_2^{\a_2}\Gamma(\a)}{\b^{\a}\Gamma(\a_1+1)\Gamma(\a_2)} \,
{}_2F_1\left(\begin{matrix}
1, \a \\
\a_1+1 
\end{matrix}\,; \frac{\b_1}{\b}\right),   \\
\dfrac{F(0)}{f(0)} = \dfrac{\a}{\b\a_1} \,
{}_2F_1\left(\begin{matrix}
1, \a \\
\a_1+1 
\end{matrix}\,; \frac{\b_1}{\b}\right), 1< \a,
\end{gathered} \quad
\quad \begin{matrix}
\a = \a_1+\a_2 \\
\b = \b_1 + \b_2
\end{matrix} 
\end{equation}

This result can be employed in theoretical considerations or as an important control or design element in practical tasks
\cite{hendrickson_centralized_2017,ranney_efficient_2021}. 
Expression \eqref{eq:F0} follows from an integral identity  \eqref{form:gammagauss} between the lower incomplete gamma $\gamma(a,x)$ function and the Gauss hypergeometric function $F(a_1,a_2;b;z)$ given by \eqref{form:gausshypgeo}. Mathematically equivalent result expressed
via the beta function can be again found as Corollary 2.1 in \citep{hendrickson_inverse_2019}.

In our time series econometrics application a $\GDD$ random variable $X$ represents distribution of different variance parameters in considered time series models. As we will see, the $\GDD$ parameters and $F(0)$ are given by the length of a time series observation and the number of time series model parameters. Then the probability of unacceptable negative variances becomes an important factor in quality diagnostics of time series models or design of time series experiments. Prior to this publication, we estimated probability $F(0)$ by very time-consuming Monte Carlo simulations \cite{gajdos_kriging_2017}.

\vspace{-6pt}

\subsubsection*{Special case: $ F(z), \a_1 = 1/2, \a_2 > 0$}
If we consider a non-integer shape parameter $\a_1 = 1/2$ as it comes out in our time series application, the lower incomplete gamma function becomes expressible \eqref{form:erf} by the error function $\textnormal{erf}(y)=(2/\sqrt{\pi}){\int_{0}^{y}e^{-z^2}dz}$
\vspace{-6pt}
\begin{equation}
\gamma\left(1/2,\b_1(x+t)\right)=\sqrt{\pi}\textnormal{erf}\left(\sqrt{\b_1(x+t)}\right)
\end{equation}
This form becomes very practical from the viewpoint of a computational implementation since the standard computation of the error function is usually much faster than computing the lower incomplete gamma function.

\vspace{-12pt}
\subsection{Statistical measures of $\GDD$}

The explicit formulae for three $\GDD$ statistical measures --- mean $\mu$, variance $\s^2$, skewness $\gamma$ are given in Klar \cite{klar_note_2015}:
\vspace{-6pt}
\begin{equation}
\mu=\frac{\a_{1}}{\b_{1}}-\frac{\a_{2}}{\b_{2}}, \quad \sigma^{2}=\frac{\a_{1}}{\b_{1}^{2}}+\frac{\a_{2}}{\b_{2}^{2}}, \quad \gamma=\frac{2\left(\a_{1} \b_{2}^{3}-\a_{2} \b_{1}^{3}\right)}{\left(\a_{1} \b_{2}^{2}+\a_{2} \b_{1}^{2}\right)^{3 / 2}}
\end{equation}

\vspace{-18pt}
\subsubsection*{Kurtosis and mode}
The kurtosis $\kappa$ and mode $\mathcal{M}$ of $\GDD$ are not explicitly considered in Klar \cite{klar_note_2015}. However, applying a general formula $\E{X^n}$ for $\GDD$ in \cite[p.~4]{klar_note_2015}, we can also derive a compact explicit expression for $\GDD$ kurtosis $\kappa \equiv \E{((X-\mu) / \sigma)^{4}}$

\vspace{-12pt}

\begin{equation}
\kappa=\displaystyle 3 + \frac{6\left(\a_1 \b_2^4 + \a_2 \b_1^4\right)}{{\left(\a_1 \b_2^2 + \a_2 \b_1^2\right)}^2},
\end{equation}
which always means positive excess kurtosis $(\kappa-3)$ or leptokurticity.

If $\a$ is less than or equal to 1, the $\GDD$ mode does not exist $(f(0)=+\infty)$. In other cases ($\a>1$), mode $\mode$ can be found by any reliable numerical optimization of scalar functions or finding a root of nonlinear equation $f'(\mode) = 0$, where
\begin{equation}\label{eq:GDD_derivativeU}
f'(z)=
\frac{\b_1^{\a_1}\b_2^{\a_2}}{\b^{\a-1}}\begin{cases} 
\dfrac{e^{z\b_2}}{\Gamma(\a_2)} \left(\begin{smallmatrix}
\b_2 U(-\a_2 + 1, -\a + 2, -\b z) - \\
- \b(\a_2-1) U(-\a_2 + 2, -\a + 3, -\b z) \end{smallmatrix}\right), & z< 0, \\[12pt]
\frac{\Gamma(\a-2)}{\Gamma(\a_1)\Gamma(\a_2)}\begin{smallmatrix}
\big(\b_2(\a_1-1)-(\a_2-1)\b_1\big)\end{smallmatrix},  & {z = 0,} \\[12pt] 
\dfrac{ e^{-z\b_1}}{\Gamma(\a_1)}\left(\begin{smallmatrix}\b(\a_1-1) U(-\a_1 + 2, -\a + 3, \b z) - \\ - \b_1 U(-\a_1 + 1, -\a + 2, \b z)\end{smallmatrix}\right), & z>0.
\end{cases}
\end{equation}

\begin{remark}\label{rem:modus}\textit{Numerical computation}
	\small
	\leavevmode\\[6pt]
	Thanks to the compactness of \eqref{eq:GDD_densU}, we obtained equation \eqref{eq:GDD_derivativeU} directly by differentiating \eqref{eq:GDD_densU} with respect to $z$ and using \eqref{form:Uab0}. This analytic form can be useful in faster computations of the mode. We found in our numerical study that computing the mode as a root of $f'(z) = 0$ can be faster than a direct scalar minimization up to two orders. 

\end{remark}	

\vspace{-12pt}
\subsection{Numerical inversion of the $\GDD$ characteristic function}

Inspired by a reference monograph \cite[ch. 3.1, 4.1]{kotz_laplace_2001} dealing with the asymmetric Laplace distribution family and its generalization\footnote{This class of probability distributions also known as Bessel $K$-function distribution or variance-gamma distribution can be regarded as a special case of $\GDD$, see Klar \cite{klar_note_2015}.}, we can also define $\GDD$ via its characteristic function. It will give us the additional fourth way of $\GDD$ computing. 

Since characteristic functions of gamma distributions $X_j$ equal to \cite{shorack_probability_2017}
\vspace{-6pt}
$$\varphi_{X_j}(t) =\left(1-\frac{it}{\b_j}\right)^{-\a_j}, t \in \Rv{}, \, j = 1,2,$$ 
using the fundamental property of the characteristic function  concerning a linear combination of independent random variables (see e.g. \cite[chap. 3]{severini_elements_2011})
\vspace{-9pt}
$$\varphi_{\Sum{i}{n}a_jX_j}(t) = \prod_{i=1}^{n}\varphi_{X_j}(a_jt),$$ 
we can write for $X = X_1-X_2 \sim \GDD(\a_1, \b_1, \a_2, \b_2)$
$$
\varphi_X(t)= \left(1-\frac{it}{\b_1}\right)^{-\a_1}\left(1+\frac{it}{\b_2}\right)^{-\a_2}.
$$
This consideration gives us an alternative, equivalent definition for $\GDD$.

\begin{definition}[\textit{\textbf{Gamma difference distribution}}]
	\leavevmode\\
	\itshape A random variable $X$ is said to have a gamma difference distribution \\ $\GDD(\a_1, \b_1, \a_2, \b_2)$ if its characteristic function is given by
	\begin{equation}
	\varphi_X(t)=  \left(1-\frac{it}{\b_1}\right)^{-\a_1}\left(1+\frac{it}{\b_2}\right)^{-\a_2}, t\in \Rv{},
	\end{equation}
	where $\a_1,\b_1,\a_2,\b_2$ are real positive constants. 
\end{definition}

The characteristic function of $\GDD$ can be easily rewritten by direct computation in the following exponential form
\vspace{-6pt}
\begin{equation}
\cf{t}=\b_1^{\a_1}\, \b_2^{\a_2}\, r(t)e^{i\phi(t)}
\end{equation} 
\vspace{-6pt}
\noindent where 
\vspace{-12pt}
\begin{align*}
r(t)&=\left(\b_1^2+t^2\right)^{-\a_1/2}\left(\b_2^2+t^2\right)^{-\a_2/2}, \\
\phi(t)&=\left(\a_1\arctan \tfrac{t}{\b_1}-\a_2\arctan \tfrac{t}{\b_2}\right).
\end{align*}
As Klar \cite{klar_note_2015} pointed out, the characteristic function and its empirical version can be applied in statistical inference for the $\GDD$, especially for parameter estimations. 

However, the characteristic function also offers us another alternative way how to effectively compute pdf and cdf of $\GDD$. This conceptually simple computational way called \textit{numerical inversion of the characteristic function}, not mentioned in \cite{klar_note_2015}, stands on the Gil-Pelaez inversion formulae \cite{gil-pelaez_note_1951} consistent with \eqref{form: Wkznuminv}
\begin{align}
f(x)=& \frac{1}{\pi} 
\int_0^\infty \Re \left[e^{-itx}\cf{t}\right]dt , \label{eq:numinvcomplfx}\\
F(x)=& \frac{1}{2}-\frac{1}{\pi} \int_0^\infty
\Im \left[e^{-itx}\cf{t}/t\right]dt. \label{eq:numinvcomplFx}
\end{align}
The method, which regularly reappears in papers during the last 60 years (see e.g. \cite{davies_numerical_1973, waller_obtaining_1995, witkovsky_numerical_2016}), seems still not widespread and usual among statisticians, data scientists, and engineers. It is worth to mention that it requires only a numerical quadrature of real functions, similarly like in previous mentioned computational ways. 

In the case of $\GDD$, Gil-Pelaez formulae together with the exponential form of $\varphi_{X}(.)$ lead to the following numerical inversion integrals for $f(x)$ and $F(x)$
\vspace{-6pt} 
\begin{align}
f(x)=&\frac{\b_1^{\a_1} \b_2^{\a_2}}{\pi}  \bigintss_{0}^{\infty} 
\frac{\cos \left(xt - \a_1\, \arctan\frac{t}{\b_1}+\a_2\, \arctan\frac{t}{\b_2}\right)}
{\left(\b_1^2+t^2\right)^{\a_1/2}\left(\b_2^2+t^2\right)^{\a_2/2}}
dt, \label{eq:nmintegralfx}\\
F(x)=&\frac{1}{2}+\frac{\b_1^{\a_1} \b_2^{\a_2}}{\pi}  
\bigintss_{0}^{\infty} 
\frac{\sin \left(xt-\a_1\, \arctan\frac{t}{\b_1}+\a_2\, \arctan\frac{t}{\b_2}\right)}
{t\left(\b_1^2+t^2\right)^{\a_1/2}\left(\b_2^2+t^2\right)^{\a_2/2}}
dt. \label{eq:nmintegralFx}
\end{align}

\begin{remark} \label{rem:splitint}\textit{Splitting integrals and the fifth-parameter $\GDD$}
	\small
	\leavevmode\\[6pt]
	Using elementary properties of complex numbers, we can also split general numerical inversion integrals \eqref{eq:numinvcomplfx}, \eqref{eq:numinvcomplFx}, and calculate them with pure oscillatory factors $\sin(xt), \cos(xt)$
	\begin{align}
	f(x)=& \frac{1}{\pi} 
	\int_0^\infty \Re\big[\cfi{t}\big] \cos(xt)dt - \frac{1}{\pi}\int_0^\infty \Im\big[\cfi{t}\big] \sin(xt)dt , \label{eq:splituminvcomplfx}\\
	F(x)=& \frac{1}{2}-\frac{1}{\pi} \int_0^\infty
	\Im\left[\tfrac{\cfi{t}}{t}\right]\cos(xt)dt+\frac{1}{\pi} \int_0^\infty
	\Re\left[\tfrac{\cfi{t}}{t}\right]\sin(xt)dt. \label{eq:splitnuminvcomplFx}
	\end{align}

\noindent In the case of more general five-parameter $\GDD$ family with the location parameter $\theta$ and the characteristic function 
	\begin{equation}
	\varphi_X(t)= e^{it\theta} \left(1-\frac{it}{\b_1}\right)^{-\a_1}\left(1+\frac{it}{\b_2}\right)^{-\a_2}, t\in \Rv{},
	\end{equation}
it is sufficient to compute all numerical inversion integrals at shifted $x \to x-\theta$.
\end{remark}

\Section{Numerical study using open data science tools}\label{sec:numstudy}

\subsection{$\GDD$ computational problem in time series kriging}

Data of many economic, financial, insurance, or business variables can be generally considered as time series datasets --– sets of observations tracking the same type of information at multiple points in time. In our econometric research, we investigate and apply a time series forecasting approach called \textit{kriging}.

The key idea of kriging consists in two stages: (1) modeling time series data using a general class of linear time series models (known as FDSLRMs), whose observations can be described by linear mixed models; (2) finding the best linear unbiased predictor (BLUP) as a prediction tool. More details from the viewpoint of theoretical framework, methodology, applications, and related computational technology can be found in our recent works \cite{gajdos_kriging_2017, hancova_estimating_2020} and our GitHub repository \textit{fdslrm} \cite{gajdos_fdslrm_2019}. 

\begin{figure}[H]
	\captionsetup{font=small}
	\centering
	\includegraphics[width=\textwidth]{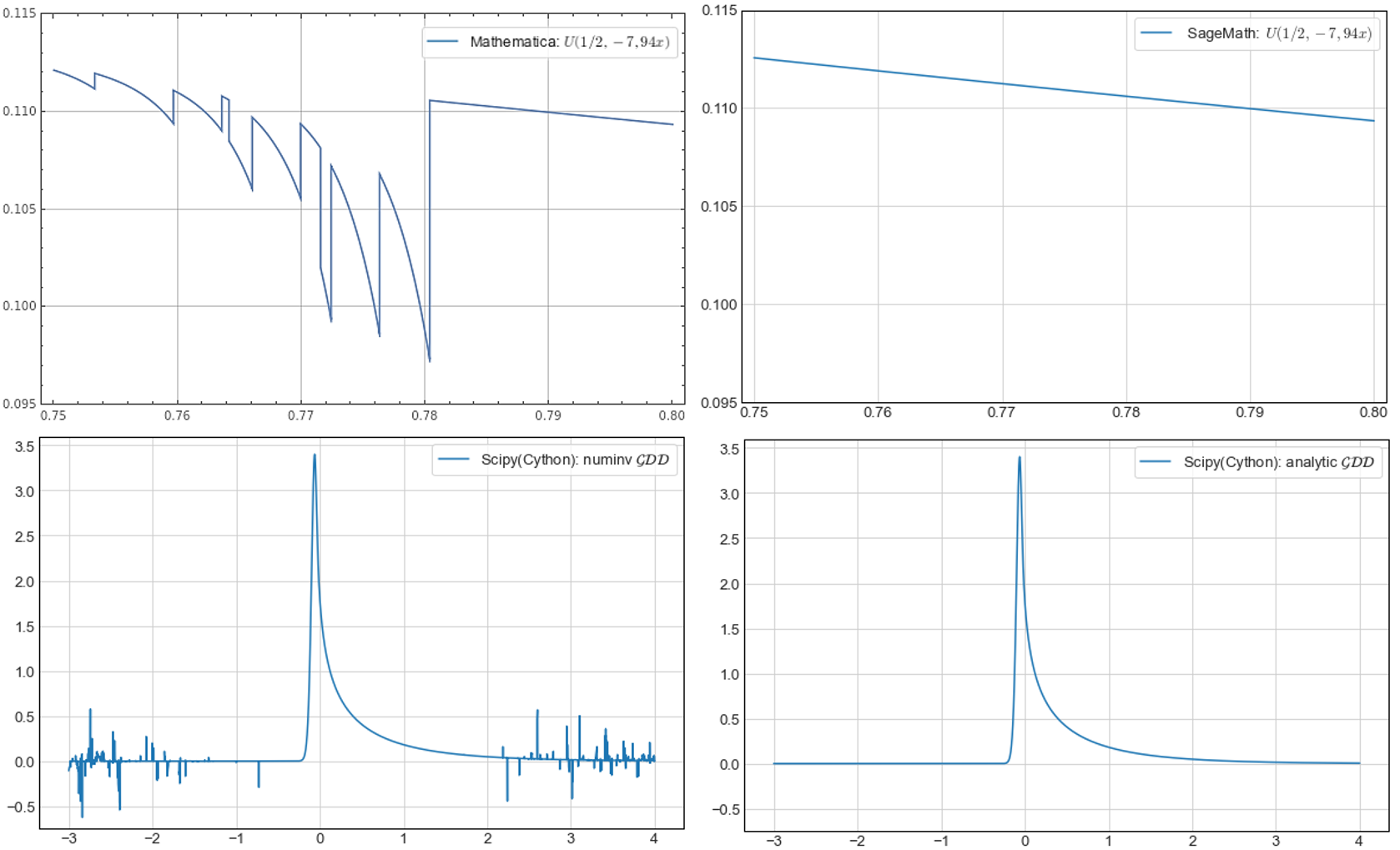}
	\caption{(1) Tricomi's hypergeometric function $U(1/2, -7, 94x), 0.75\leqq x\leqq 0.80$ \\ generated by Mathematica 12.3 (\textit{upper left}) and by SageMath 9.2 (\textit{upper right}). \\ (2) Pdf for $\GDD$ from a standard numerical quadrature of the numerical inversion integral (\textit{lower left}) and from the analytic formula  (\textit{lower right}) in SciPy(Cython).}
	\label{fig:GDDproblem}
\end{figure}

To calculate and explore properties of BLUP in real data analysis or simulations, we need to estimate variance parameters of a chosen time series model. As we can see in appendix B, the well-known method of moments leads to unbiased estimators with $\GDD$ distribution.
In our case of particular time series dataset (electricity consumption), we deal with the following typical values of $\GDD$ parameters \eqref{eq:GDDtypical}
\begin{equation}\label{eq:GDDvalues}
\begin{gathered}
\a_1 = 0.5, \a_2 = 8.5, \,\, \b_1 = 1.0, \b_2 = 93.0, \\
\a=\a_1+\a_2 = 9.0, \,\,\b = \b_1+\b_2 =94.0. \\
\end{gathered}
\end{equation}
If we try to calculate or plot $\GDD$ pdf $f(x)$ on interval $(-3,4)$, we get very inconsistent results as for computing accuracy, reliability, or speed (fig. \ref{fig:GDDproblem}).

\newpage
\begin{remark}\label{rem:SixSigma}\textit{Six-sigma rule for $\GDD$}
	\small
	\leavevmode\\[6pt]
Regarding obtained $\GDD$ parameters \eqref{eq:GDDvalues}, such $\GDD$ has $\mu = 0.41, \sigma^2 = 0.50, \gamma = 2.8, \kappa= 15$ and  $ \mode = -0.062$. Due to $\GDD$ asymmetry, we chose interval $(-3,4)$ in accordance with \textit{six-sigma rule}. The rule gives us exactly $(-3.84, 4.66)$, which we cut to $(-3,4)$ for simplicity, still having a high coverage of $X$ described by probability $P(X \in (-3,4)) = 0.996$.
\end{remark}

For the first time, we encountered these computational problems in Mathematica 11, when computations of analytic expression \eqref{eq:GDD_densU} using $U(1-\a_1, 2-\a,x\b) = U(1/2, -7, 94x)$ completely failed in some points of the interval\footnote{see details in our online Jupyter notebook  \href{https://nbviewer.jupyter.org/urls/dl.dropbox.com/s/fes4805h0kqy24a/TricomiUfunction_Sage83_Mathematica11.ipynb}{\underline{Problem -- Mathematica 11}}}. As it is depicted in fig. \ref{fig:GDDproblem} (upper left), the problem still hasn't disappeared\footnote{see details in our online Jupyter notebook \href{https://nbviewer.jupyter.org/urls/dl.dropbox.com/s/vslw2vwnkyh1f61/TricomiUfunction_Sage92_Mathematica12_3.ipynb}{\underline{Problem -- Mathematica 12}}} in the latest Mathematica 12.3 \cite{wolfram}. Later we found a paper \cite{johansson_computing_2019} describing general problems in $U(a,b,z)$ implementation. We will examine this Mathematica problem, which e.g. does not appear in open CAS software SageMath (see fig. \ref{fig:GDDproblem}, upper right), in our numerical study.

Such potential problems with \eqref{eq:GDD_densW} are also pointed by Klar  \cite{klar_note_2015}, who therefore recommends that maybe numerical quadrature of the pdf convolution integral \eqref{eq:GDDdensconv} is more convenient. But in the case of our $\GDD$, we run into problems with both integral versions of pdf --- convolution integral \eqref{eq:GDDdensconv} and numerical inversion integral \eqref{eq:nmintegralfx}.

If we use automatic built-in integrators for numerical quadrature of \eqref{eq:GDDdensconv}, then the accuracy of results is much better, e.g. both commercial Mathematica and open SciPy \cite{scipy} generate the same plot in fig. \ref{fig:GDDproblem}(lower right). But runtimes are much worse, not reasonable in the frame of any computational research. 

On the other hand, built-in numerical quadratures of the numerical inversion integrals give us very unreliable values on pdf tails (see example in fig. \ref{fig:GDDproblem}, lower left) and lead to various warnings in the integrators, e.g. slow convergence; probably divergent integral; the result may be incorrect or use a special-purpose integrator. If we use recommended specialized integrators for oscillatory integrands, results are again much better, also for pdf tails, but runtimes become very unfavorable (much worse than in the case of the convolution integral)\footnote{e.g. runtime for a plot of \eqref{eq:nmintegralfx} with 200 points equals $2.6$~\!s in Scipy and $1.6$~\!s in Mathematica.}. 

These inconsistent results convinced us to carry out a systematic numerical study that would reveal the weaknesses and benefits of considered computing methods in various open and commercial digital tools.

\subsection{Digital tools and numerical quadratures}

\subsubsection*{Open vs. commercial computing tools}
During the last ten years, free open software and data science tools, based on programming languages Python \cite{python3} and R \cite{Rcore}, together with interactive environments for easily shareable, modifiable, reproducible, and collaborative work --- especially Jupyter \cite{kluyver_jupyter_2016} and RStudio \cite{rstudio}, have conquered the data science world \cite{frederickson_ranking_2019,kaggle_kaggles_2020} and have become available to everyone. In appendix C we summarize all open digital tools with their versions, which we applied in our numerical study. 

Our main experimental tools were Sage and Python, but for benchmark purposes we also compared results with two major commercial scientific computing systems --- Wolfram Mathematica \cite[v.~12.3, abbr.~\MMA]{wolfram},  and MATLAB \cite[v.~9.10, abbr.~\MTB]{matlab}.
To test some free \MTB~code, it was also convenient for us to use Octave, the open source \MTB~clone (see box C1 in appendix C). 

\subsubsection*{Computation conditions and hardware specifications}

All computations in given digital tools were run on a Windows 10 (64-bit) laptop equipped with an Intel i7-9850H CPU @ 2.60 GHz (6 cores) and 64 GB RAM. The Python, Sage and \MMA~were installed from their official repositories. \MTB~computations were performed using a trial \MTB~Windows (64-bit) version (R2021a).

\vspace{-6pt}
\subsubsection*{Built-in functions and numerical integrators}

We used arbitrary-precision library PARI/GP to generate quickly high-precision (quadruple 128-bit precision) values of $f(x)$ using \eqref{eq:GDD_densU}. The results were mutually cross-checked using arbitrary-precision mpmath and Arb\footnote{According to \cite{johansson_computing_2019}, arbitrary-precision Arb is probably the best in rigorous computing $U(a,b,z)$.}. Then we can check the accuracy of analytic expression \eqref{eq:GDD_densU} in \MMA, Sage, SciPy, and R, which have $U(a,b,z)$ implementation. In the case of CAS software, \MMA~and Sage, we can work symbolically and then express final results numerically by appropriate commands
or directly work numerically in machine precision mode (53-bit) as it is in SciPy and R. Regarding numerical quadrature of the convolution integral or numerical inversion integrals, \MMA, Sage, SciPy, and R have built-in automatic integrators dominantly applying a standard workhorse of numerical integration -- Gauss quadrature and its various forms. But we can also find specialized integrators for specific types of integrands.

\begin{remark}\label{rem:commands}\textit{Specific commands}
	\small
	\leavevmode\\[6pt]
As for numerical approximation of symbolic results, in \MMA~we use command \texttt{N[ ]} and in Sage we have two alternatives:~\texttt{.n()} or \texttt{fast\_float(...)} for fast numerical evaluation of functions. Speaking about Tricomi's $U(a,b,z)$, SciPy has not only a standard implementation in library \texttt{scipy.special}, but also a fast Cython version in library \texttt{scipy.special.cython\_special}. In the case of R, $U$ implementation can be found e.g. in package \texttt{fAsianOptions}.

Numerical integration can be realized in Sage by \verb|numerical_integral| command, which calls GSL library; in SciPy \texttt{quad} and in R \texttt{integrate}, both using a technique from the Fortran library QUADPACK; in \MMA~command \texttt{NIntegrate[]}. Moreover, both Sage and SciPy allow programming fast Cython versions of integrands.
We can also choose special integrators, e.g. in the case of \MMA\footnote{\url{https://reference.wolfram.com/language/tutorial/NIntegrateIntegrationStrategies}}, mpmath or Sage (via GSL library), we can also find oscillatory integrators (e.g. Logmann, Clenshaw-Curtis, or double exponential oscillatory method) more convenient for our Fourier type numerical inversion integrals.
\end{remark}

\vspace{-12pt}
\subsubsection*{Trapezoidal rule and DE quadrature} 

In general, $\GDD$ convolution or numerical inversion integrals can be computed by any suitable numerical quadrature method. Here we briefly introduce the so-called DE quadrature, which is a conceptually simple, very efficient, and highly precise method, but still not very familiar and fully appreciated in statistical computing\footnote{The DE formula is successfully used e.g. in molecular physics, fluid dynamics, civil and financial engineering. It is useful for evaluation of indefinite integrals, for solving integral or differential equations 	\cite[see e.g.][]{mori_discovery_2005, lovrod_double_2019}.}.

The DE quadrature deals with the well-known trapezoidal rule, which despite its simplicity, may be very efficient --- exponentially fast and accurate for certain types of integrals over the real line (see details in \cite[ch.~5.4, 5.5]{gil_numerical_2007}, \cite{trefethen_exponentially_2014}). The trapezoidal rule  computes such integral of a real function $w(x)$  approximately as ($h >0, n \in \mathbb{N}$) 
\vspace{-6pt}
\begin{equation}\label{eq:trruleinf}
\displaystyle I(w)=\int_{-\infty}^{\infty}w(x) dx \approx h \sum_{k=-n}^{n} w(x_k) = h \sum_{k=-n}^{n} w(kh). 
\end{equation}

\vspace{-6pt}

\noindent Considering this efficiency, it is natural to ask whether other types of integrals can be transformed to a form leading to exponentially fast and accurate trapezoidal rule, e.g. by changing the variable of integration
\vspace{-6pt}
\begin{equation}\label{eq:TRvartrans}
\begin{gathered}
\displaystyle I(w)=\bigintsss_a^{b}w(x) dx \quad a,b \in \mathcal {R} \cup \left\{-\infty ,+\infty \right\}, 
\\[6pt]
\displaystyle I(w) \underset{x\,=\, \Phi(t)}{=} \bigintsss_{-\infty}^{\infty} w\Big(\Phi(t)\Big) \Phi'(t) d t \approx h \sum_{k=-n}^{n} w\Big(\Phi(k h)\Big) \Phi'(k h).
\end{gathered}
\end{equation}

Such variable transformation exists, and it was discovered and further developed for several general types of integrals by Japanese mathematician Mori and his collaborators \cite{takahasi_double_1974, mori_discovery_2005}. If we employ a function $x = \Phi(t), a=\Phi(-\infty),  b=\Phi(\infty)$, such that the transformed integrand in \eqref{eq:TRvartrans} decays as a double exponential function ($c \in \R$)

\vspace{-18pt}

 \begin{equation}\label{eq:DEcond}
 \left|w\Big(\Phi(t)\Big) \Phi'(t)\right| \rightarrow \exp\left(-c\cdot e^{|t|}\right) \text{ for }\, t \rightarrow \pm \infty,
 \end{equation}
 
\vspace{-6pt}

\noindent than we get the exponentially convergent trapezoidal rule \eqref{eq:TRvartrans} whose error behaves as $\bigO \big(\exp(-CN/\ln N )\big), N =2n+1, C \in \R$.
The trapezoidal rule with the double exponential (DE) transformation is called \textit{the double exponential quadrature} (DE quadrature). Mori et al \cite{mori_discovery_2005, sugihara_optimality_1997} also proved that under very general conditions and sufficiently big $N$ such quadrature formula surprisingly appears optimal in the sense that there does not exist any other quadrature formula obtained by variable transformation whose error decays faster.

Particular transformations of the DE quadrature which satisfied \eqref{eq:DEcond} can be found e.g. in \cite{mori_double-exponential_2001} and two of them, which are directly connected to our convolution \eqref{eq:GDDdensconv}  and numerical inversion integrals \eqref{eq:nmintegralfx}, \eqref{eq:nmintegralFx},  are shown in tab. \ref{tab:DEtrans} (for the integral with $\cos(\omega x)$ it is sufficient to shift $t \to t-1/(2\pi)$ in $\phi(t)$).

\begin{table}[H]
	\begingroup
	\renewcommand{\arraystretch}{2.5}
	\begin{center}
	\caption{Double exponential transformations for two types of integrals.}	\label{tab:DEtrans}
	\begin{tabular}{r| l}
			\hline \hline \\[-30pt]
			\small
			$\displaystyle \bigintsss_{a}^{\infty} w(x)e^{-bx} d x$  & $\begin{aligned}
			x & =a+b^{-1}\phi(t) \\[3pt]
			\phi(t) & = \exp(t-\exp(-t))
			\end{aligned} $  \\[12pt]
			\hline \\[-24pt]
			$\displaystyle\bigintsss_{a}^{\infty} w(x) \sin(\omega x) \, dx$ & 	$\begin{aligned}
			x &= a+M\omega^{-1}\phi(t) \\[3pt]
			\phi(t)  & =
			\frac{t}{1-\exp \Big(-2 t-\alpha\left(1-\mathrm{e}^{-t}\right)-\beta\left(\mathrm{e}^{t}-1\right)\Big)} \\[3pt]
			\footnotesize
			\alpha & =\frac{1}{4\sqrt{1+\frac{1}{4\pi}M \ln (1+M)}}, \quad Mh =\pi
			\end{aligned}$    \\[40pt]			
			\hline \hline
		\end{tabular}
	\end{center}
	\endgroup
\end{table}

\begin{remark}\label{rem:opencode}\textit{Open source code}
	\small
	\leavevmode\\[6pt]
We found two open source computational implementations of the trapezoidal rule and DE quadrature that became very suitable for our numerical study.
\newpage
\begin{itemize}
	\setlength\itemsep{3pt}
	\item \MTB~package \texttt{CharFunTool} of Witkovsky \cite{witkovsky_numerical_2016} implementing the simple trapezoidal rule over six-sigma interval for numerical inversion of any characteristic function. Using Octave as a testing tool, we rewrote its code \cite{witkovsky_witkovskycharfuntool_2021} with minimal modification into R, Python, and NumPy.
	\item Fortran and C implementation of DE quadrature\footnote{Implementation includes also  cases $\int_a^{\infty} f(x)dx $ for non-oscillatory and oscillatory integrand $f(x)$.} from Ooura, whose with Mori developed the DE oscillatory transformation \cite{ooura_robust_1999,ooura_double_2005} shown in the second row in tab. \ref{tab:DEtrans}. Thanks to the conceptual simplicity and great readability of Ooura's code \cite{ooura_oouras_2006}, we were able to rewrite it into R, Python, and Numba with generalization from \texttt{f(x)} to \texttt{f(x,*params)}, an integrand with a variable number of extra parameters. 
\end{itemize}
\noindent Our open codes and Jupyter notebooks from the entire numerical study are available at our GitHub repository \cite{gajdos_fdslrm_2019,gajdos_fdslrmgdd_2021}.
\end{remark}

\vspace{-18pt}
\subsection{Numerical study results and discussion}
\vspace{-6pt}
\subsubsection*{Numerical results}
We realized one numerical experiment for each computational method and available digital tool. Such experiment involved 3 runs, and each run contained 10 realizations of computing $\GDD$ at $10~000$ points uniformly distributed on six-sigma interval $(-3,4)$ for our time series kriging $\GDD$ parameters \eqref{eq:GDDvalues}. It means that one numerical experiment performed $3\x10^5$ calculations. We used built-in commands or functions for measuring the execution time of each numerical experiment\footnote{The command \texttt{\%timeit -r 3 -n 10} is available in all Python-based tools. In the case of \MMA~we applied command \texttt{RepeatedTiming[.]}. In \MTB, Octave and R we used a system time function.}.

\begin{figure}[H]
	\captionsetup{font=small}
	\centering
	\includegraphics[width=0.8\textwidth]{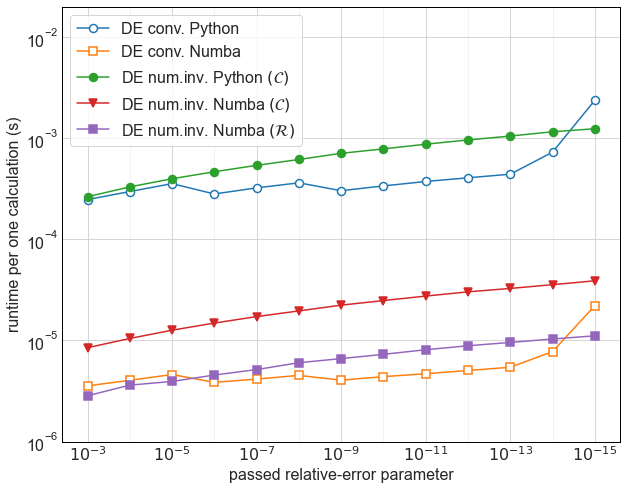}
	\caption{Average DE runtime per calculation for the $\GDD$ pdf  \textit{versus} a passed input relative-error parameter, using a standard \textit{Python} and its high-performance compiler \textit{Numba}.  \\ (Each point represents the summary result from one numerical experiment.)}
	\label{fig:DEruntime}
\end{figure}

\begin{figure}[H]
	\captionsetup{font=small}
	\centering
	\includegraphics[width=0.8\textwidth]{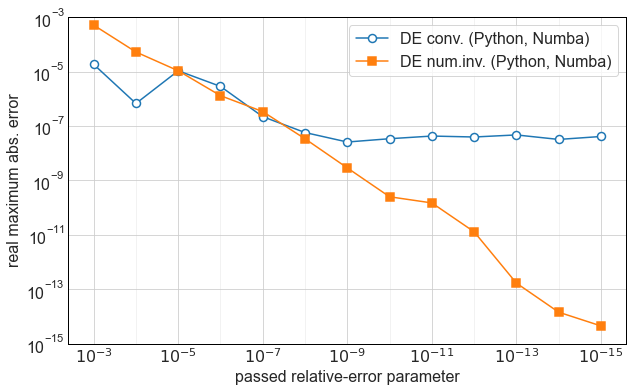}
	\caption{Real accuracy (maximum abs. error) in the DE quadrature for the $\GDD$ pdf \\ given by the convolution integral \eqref{eq:GDDdensconv} and the numerical inversion integrals  (\ref{eq:nmintegralfx}, \ref{eq:numinvcomplfx}) \\ \textit{with respect to }a passed input relative-error parameter $\epsr$, using \textit{Python} and \textit{Numba}. \\(Each point represents the summary result from one numerical experiment.)}
	\label{fig:DEerror}
\end{figure}

The benchmark results from one numerical experiment contain average runtime, runtime standard deviation (which was less than 5\% of runtimes for all realizations in all experiments), acceleration (compared to the Python implemented trapezoidal rule), and accuracy (real maximum absolute error). The entire numerical study is represented by 90 numerical experiments, whose results are summarized by tab.~\ref{tab:numstudy}, fig.~\ref{fig:DEruntime}, and fig.~\ref{fig:DEerror}. 

Particularly, the benchmark for given methods and tools includes
\begin{itemize}
	\setlength\itemsep{3pt}
	\item \textit{18~experiments} (tab.~\ref{tab:numstudy} -- first two blocks) using a built-in quadrature (default precision) of convolution integral \eqref{eq:GDDdensconv} or using analytic expression \eqref{eq:GDD_densU}:  \MMA($1\xm$symb. calculation, $2\xm$53-bit num. calculation), Sage($2\xm$symb., $4\xm$53-bit), SciPy($4\xm$53-bit), R($2\xm$53-bit), mpmath($1\xm$53-bit), PARI/GP($2\xm$\texttt{p20}, \texttt{p15}), 
	\item \textit{6~experiments} (tab.~\ref{tab:numstudy} -- third block) using the trapezoidal rule: \MTB($1\xm$53-bit), Octave($1\xm$53-bit), R($2\xm$53-bit), Python($1\xm$53-bit), NumPy($1\xm$53-bit)
	\item \textit{26~experiments} (fig.~\ref{fig:DEruntime}, fig.~\ref{fig:DEerror}, tab.~\ref{tab:numstudy} -- fourth block) using the DE quadrature of convolution integral \eqref{eq:GDDdensconv}:  Python ($13\xm$53-bit, $\epsr = 10^{-3}, 10^{-4},\ldots, 10^{-15}$) and Python with Numba ($13\xm$53-bit, $\epsr = 10^{-3}, \ldots, 10^{-15}$) 
	\item \textit{39~experiments} (fig.~\ref{fig:DEruntime}, fig.~\ref{fig:DEerror}, tab.~\ref{tab:numstudy} -- fourth block) using the DE quadrature of the numerical inversion integral in $\C$ \eqref{eq:numinvcomplfx} or directly in $\R$ \eqref{eq:nmintegralfx}:  Python ($13\xm$53-bit --- in $\C$, $\epsr = 10^{-3}, \ldots, 10^{-15}$), Python with Numba ($26\xm$53-bit --- in $\C$ and in $R$, $\epsr = 10^{-3},\ldots, 10^{-15}$)
	\item \textit{1~experiment} (tab.~\ref{tab:numstudy} -- fourth block) using the DE quadrature of the numerical inversion integral \eqref{eq:GDDdensconv} in R(53-bit, $\epsr = 10^{-3}$)
\end{itemize}

\begin{table}[H] 
\renewcommand{\arraystretch}{1.25}
\captionsetup{font=small}
\caption{Average runtimes, accelerations (with respect to the trapezoidal rule for (\ref{eq:numinvcomplfx}) implemented in Python with complex numbers), and real accuracy (maximum abs. error) \\ for $\GDD$ pdf calculations using analytic and numerical methods in various digital tools. \\(Each row represents the summary results from one numerical experiment.)\vspace{-12pt}}	\label{tab:numstudy}
\catcode`\-=12
\begin{center}
\small
\begin{tabular}{clccc}
\hline\hline \\[-9pt] \multicolumn{5}{c} { prob. density function $f(x)$ for $\GDD$  (calculating $10\,000$ points over an interval) } \\[3pt] \hline \\[-9pt]
method & digital tool & run time (s) & acceleration & accuracy \\[3pt] \hline \hline\\[-9pt] 

\multirow{7}{*}{\footnotesize$\left[\begin{array}{c}\text{built-in} \\ \text{numerical} \\ \text{integration } \\  \text{of } \eqref{eq:GDDdensconv}, \\ \text{ default } \\ \text{precision }	\end{array} \right]$ } 
&\underline{\textit{MMA}$^*$(\textit{53-bit})	              }& $\mathit{39.1}$ & $\mathit{0.05}$ & $\mathit{5 \!\x\! 10^{-8\phantom{0}}}$  \\ 
&Sage (53-bit) 	                  & $124\phantom{.}$ & $0.02$ & $2 \!\x\! 10^{-7\phantom{0}}$ \\
&Sage (\verb|fast_float|, 53-bit) & $0.81$ & $2.30$ & $2 \!\x\! 10^{-7\phantom{0}}$ \\
&Sage (Cython, 53-bit)			  & $1.02$ & $1.82$ & $2 \!\x\! 10^{-7\phantom{0}}$\\
&SciPy (53-bit)				  & $66.7$ & $0.03$ & $7 \!\x\! 10^{-9\phantom{0}}$\\
&SciPy (Cython, 53-bit)		  & $1.16$ & $1.60$ & $7 \!\x\! 10^{-9\phantom{0}}$\\
&R (53-bit)		  & --- & --- & ---\\[6pt]
\hline\hline \\[-6pt] 

\multirow{12}{*}{\footnotesize$\left[\begin{array}{c} \text{analytic} \\ \text{expression}  \\  \text{for } f(x) \\ (\ref{eq:GDD_densU}) \\   \text{with } \\ U(a,b,z)	\end{array} \right]$ } 
&\underline{\textit{MMA}$^*($\ttfamily{N[ ]}$,\mi{53}$-$\mi{bit})$}  & $\mi{9.09}$ & $\mi{0.20}$ & $\mi{5 \!\x\! 10^{-15}}$   \\
&Sage (\verb|.n()|, 53-bit) 	   & $8.62$ & $0.22$ & $3 \!\x\! 10^{-15}$   \\
&Sage (\verb|fast_float|, 53-bit)  & $3.78$ & $0.49$ & $3 \!\x\! 10^{-15}$ \\[6pt]
\cline{2-5} \\[-6pt]
& mpmath (53-bit) 				   & $4.09$ & $0.45$ & $3 \!\x\! 10^{-15}$ \\
& PARI/GP (\verb|p20|, 128-bit) 	   & $1.16$ & $1.60$ & $2\!\x\!10^{-16}$ \\
& PARI/GP (\verb|p15|, 64-bit) 	   & $0.32$ & $5.80$ & $5\!\x\!10^{-14}$ \\[9pt] \cline{2-5} \\[-6pt]

&\underline{\textit{MMA}$^{\dagger}(\mi{53}$-$\mi{bit})$} 	& $\mi{2.02}$ & $\mi{0.92}$ & $\mi{3 \!\x\! 10^{-2\phantom{0}}}$   \\
& Sage (53-bit) 	& $3.95$ & $0.47$ & $3 \!\x\! 10^{-15}$ \\

& SciPy (53-bit)				& $0.40$ & $4.65$ & $5 \!\x\! 10^{-10}$\\
& SciPy ({Cython}, 53-bit)				& $0.06$ & $32.2$ & $5 \!\x\! 10^{-10}$ \\
&R (53-bit)		  & --- & --- & ---\\[6pt]

\hline\hline \\[-6pt] 

\multirow{6}{*}{\footnotesize$\left[\begin{array}{c} \text{trapezoidal} \\ \text{ rule for} \\ (\ref{eq:numinvcomplfx}) 
\end{array} \right] $  } 
&	\underline{\textit{MTB}$^{\mathsection} (\mathcal{C},$ \ttfamily{tol} $\!\!=\!10^{-4})$} & {$\mi{0.13}$}&  {$\mi{14.3}$} & {$\mi{9 \!\x\! 10^{-4\phantom{0}}}$}  \\
&	Octave $(\mathcal{C},$ \ttfamily{tol} $\!\!=\!10^{-4})$ & {$0.54$}&  {$3.40$} & {$9 \!\x\! 10^{-4\phantom{0}}$}  \\
&	R $(\mathcal{C},$ \ttfamily{tol} $\!\!=\!10^{-4})$ & {$2.94$}&  {$0.63$} & {$9 \!\x\! 10^{-4\phantom{0}}$}  \\
&	R $(\text{vec}, \mathcal{C},$ \ttfamily{tol} $\!\!=\!10^{-4})$ & {$0.37$}&  {$5.08$} & {$9 \!\x\! 10^{-4\phantom{0}}$}  \\
&	{\textbf{Python} $\boldsymbol{(\mathcal{C},}$ \ttfamily{tol} $\boldsymbol{\!\!=\!10^{-4})}$}
& {$\boldsymbol{1.85}$}&   {$\boldsymbol{1.00}$} & {$\boldsymbol{9 \!\x\! 10^{-4\phantom{0}}}$} \\
&	NumPy $(\mathcal{C},$ \ttfamily{tol} $\!\!=\!10^{-4})$ & {$0.28$}&  {$6.58$} & {$9 \!\x\! 10^{-4\phantom{0}}$} \\[6pt]
\hline\hline \\[-6pt]

\multirow{6}{*}{\footnotesize$\left[\begin{array}{c} \text{trapezoidal}  \\ \text{rule with} \\ \text{ the DE} \\ 
\text{transformation} \\ \text{for } (\ref{eq:numinvcomplfx}) \text{ in } \mathcal{C}  \\ \text{or } (\ref{eq:nmintegralfx}) 
\text{ in } \mathcal{R} \end{array}\right] $ }  
&	R $(\mathcal{C}, \epsr \!=\!10^{-3})$ & {$1.49$}&  {$1.24$} & {$5 \!\x\! 10^{-4\phantom{0}}$}  \\
& Python $(\mathcal{C}, \epsr \!=\!10^{-3})$&  $2.63$  & $0.70$ & $5 \!\x\! 10^{-4\phantom{0}}$\\
& Numba $(\mathcal{C},\epsr \!=\!10^{-15})$  &  $0.38$  & $4.87$ &  $4 \!\x\! 10^{-15}$ \\
& Numba $(\mathcal{R}, \epsr \!=\!10^{-15})$ &  $0.11$  & $16.9$ & $4 \!\x\! 10^{-15}$ \\
& Numba $(\mathcal{R}, \epsr \!=\!10^{-10})$ &  $0.07$  & $26.0$ &  $3 \!\x\! 10^{-10}$ \\
& Numba $(\mathcal{R}, \epsr \!=\!10^{-3})$  &  $0.03$  & $68.7$ & $5 \!\x\! 10^{-4\phantom{0}}$ \\[6pt]
	\hline\hline
\multicolumn{5}{l}{$^{\phantom{*}*}$\footnotesize \textit{MMA} $\equiv$ Wolfram Mathematica \cite[v.~12.3]{wolfram}} \\[-3pt]
\multicolumn{5}{l}{$\,\,^{\dagger}$\footnotesize \textit{MMA}'s problematic implementation of  $U(a,b,z)$ in MachinePrecision (53-bit) mode (\href{https://nbviewer.jupyter.org/urls/dl.dropbox.com/s/vslw2vwnkyh1f61/TricomiUfunction_Sage92_Mathematica12_3.ipynb}{\underline{link}})} \\[-3pt]
\multicolumn{5}{l}{$\,\,^{\mathsection}$\footnotesize \textit{MTB }$\equiv$ MATLAB \cite[v.~9.10]{matlab}; computations were realized using an \textit{MTB} trial version}
\end{tabular}
\end{center}
\end{table}

\phantom{x}
\vspace{0.5cm}

\subsubsection*{The pdf convolution integral and analytic expression with $U(a,b,c)$}
The worst results were obtained by R that totally failed in a built-in integration of the convolution integral and in calculating the analytic expression (the result of the package was a completely different function). It confirms the fact that R is more suitable for data processing and for applying statistical models and procedures \cite{chambers_breaking_2017} than to analytic or numerical tasks where R should be used with caution.

As for CAS software -- \MMA~and Sage, the results are comparable with two exceptions. The first one deals with the accuracy of \MMA~running in MachinePrecision mode (53-bit), where computing pdf analytically leads to damaged results (low accuracy $\approx 10^{-2}$)\footnote{The \MMA~problem disappears if before calculations we convert argument $z$ to a closest rational approximation (command \texttt{Rationalize[.]}) and then we apply numerical 53-bit approximation (with command \texttt{N[.]}). But this extra step requires about 4.5 times longer runtime.}.
The second exception is connected to Sage. Using command \texttt{fast\_float}, which creates a fast-callable version of given functions, we get the fastest CAS results of the analytic expression with double 53-bit precision. It also accelerates a built-in numerical quadrature of the convolution integral to speeds of cythonized SciPy.

Concerning arbitrary-precision and numerical libraries, mpmath has a speed of CAS systems. One-order faster PARI and SciPy are comparable with respect to speeds, but PARI has no problem to achieve the double-precision\footnote{PARI does not allow us to set exactly the double 53-bit precision. You can set 64-bit or 128-bit as closest.}. The only speed exception is SciPy with cythonized $U(a,b,z)$, where runtimes are hundreds of second --- approximately one-order faster than PARI and SciPy and two-order faster than CAS systems.

\vspace{-6pt}
\subsubsection*{Trapezoidal rule and DE quadrature}
The pure R and Python versions of Witkovsky's trapezoidal rule implementation without vectorization has at least  twice the speed of CAS software (with accuracy $\approx 10^{-15}$), but with much lower accuracy $\approx 10^{-4}$, which is the default precision of the implementation. Vectorization in R and Python (using NumPy) accelerates computing up to one order to speeds of \MTB.

The DE oscillatory quadrature for the numerical inversion integral written in pure R and Python (without vectorization) is comparable with the trapezoidal rule alone. Using high-performance Python compiler Numba, we get the following accelerations with respect to other computational tools and methods
\begin{itemize}
	\small
	\setlength{\itemindent}{-6pt}
	\setlength\itemsep{-9pt}
	\item accuracy $\approx 10^{-15}$: $85\xm$faster than \MMA(\texttt{N[]}), \\
	\phantom{xxxxxxxxxxxxxx}	   $35\!-\!40\xm$faster than Sage(\texttt{fast\_float} or 53-bit) \& mpmath, \\
	\phantom{xxxxxxxxxxxxxx}  $5\!-\!10\xm$faster than PARI, \\
	\item accuracy $\approx 10^{-10}$: $6\xm$faster than SciPy, \\
	\phantom{xxxxxxxxxxxxxx}	$0.8\xm$slower than Cython,\\
	\item accuracy $\approx 10^{-4\phantom{0}}$: $75\xm$faster than \MMA(53-bit) \\
	\phantom{xxxxxxxxxxxxxx} $70\xm$faster than Python, \\
	\phantom{xxxxxxxxxxxxxx} $10\xm$faster than NumPy \& R, \\ 
    \phantom{xxxxxxxxxxxxxx}	$5\xm$faster than \MTB
\end{itemize}

We also tested the DE non-oscillatory quadrature in the case of the convolution integral. Its results, in comparison with DE oscillatory version for the numerical inversion integral, were summarized in fig.\ref{fig:DEruntime} (average runtime per calculation $f(x)$ at one point) and in fig.\ref{fig:DEerror} (real maximum error with respect to a passed input relative-error parameter). We can see that they are comparable in speeds (fig.\ref{fig:DEruntime}). From the accuracy viewpoint, they are equivalent for input parameter $\epsr \geq 10^{-8}$, but for lower $\epsr$ the non-oscillatory version is stacked (due to achieving predefined maximum nodes), whereas the oscillatory version consistently increases its accuracy.

Finally, we mention that $f(x)$ can be successfully calculated as a numerical derivative of cdf \eqref{eq:GDD_DF}, but in this case we lost accuracy or increase significantly runtimes.

\section{Conclusions}

In this paper we revisited and extended in some details the existing theoretical framework for statistical computing of the gamma difference distribution $(\GDD)$. In connection with our theoretical considerations, we also explored the practical aspects of four different computational ways  and corresponding computing tools for $\GDD$ with special attention to open software based on R and Python. 

Using open Python-based data science tools, we have created our own tool for statistical computing of $\GDD$. It combines numerical inversion of $\GDD$ characteristic function with the double exponential (DE) oscillatory quadrature, on the basis of the original C package from Ooura --- one of the DE quadrature originators. By the numerical study, including $N = 90$ numerical experiments (four different computational approaches, 10 open and 2 commercial computing tools), we have demonstrated that our open code implemented in high-performance Python(with Numba) became the best in all aspects --- speed, high precision, and reliability.

Specifically, our tool has reached speeds of Cython, whose performance is typical of the order of C programming language. Simultaneously it can calculate $\GDD$ with high precision. At the double 53-bit precision, it outperformed the speed of the analytical computation based on Tricomi's $U(a,b,z)$ function in CAS software (commercial Mathematica and SageMath) by 1.5-2 orders. At the precision of scientific numerical computing tools, it exceeded open SciPy, NumPy and commercial MATLAB 5-10 times. It is also worth to mention that the speed of our tool is not final. We assume that parallelization will lead to at least one order faster code \cite{boulle_high-performance_2019} than now. This is our current work in progress since the need to keep track of internal states in some loops of the DE quadrature makes it difficult to find an optimal parallelization approach. 

However, these conclusions are not yet generally valid since our numerical study was performed only with $\GDD$ parameters which resulted from
our $\GDD$ application in time series kriging --- a forecasting approach using linear mixed models with the best linear unbiased prediction \cite{gajdos_fdslrm_2019,hancova_estimating_2020}. Our preliminary results with other parameters and functions indicate that the tool could be effective in much wider conditions, but it must be proved by a more complex numerical study which is under our current intensive investigation. 
After successful confirmation, our tool could be useful also for R statistical community. Although our R implementation is 1-1.5 order slower, it could be appreciated by R users since our numerical study showed that all currently available tools in R failed in calculations with our $\GDD$ parameters. 

The fast and effective general implementation of the DE quadrature in numerical inversion of a mixture of characteristic functions could open some new possibilities for data analysis based on exact probability distributions (not on asymptotic or approximate methods that often require large data or a lot of runtime to be applied correctly) for areas like multidimensional statistical data analysis, measurement uncertainty analysis in metrology as well as in financial mathematics and risk analysis. 

We are fully aware that there are also other numerical methods and digital tools which could be used in our calculations. For example, there is a highly efficient implementation of DE quadrature in Julia \cite{slevinsky_use_2015,slevinsky_mikaelslevinskydequadraturejl_2020}; or the sinc function transformation could be used as an alternative to DE transformation \cite{stenger_handbook_2010}; or we could probably use the complex quadrature for our integrals \cite{asheim_complex_2013, deano_computing_2017}. But these methods and their implementations required a much more sophisticated approach than our DE quadrature. 

Finally, we would like to highlight the general benefits of open data science tools. Thanks to open data science, we could connect the plethora of computing tools together into a coherent framework for the numerical study. Availability, transparency, comprehensibility and effective customization resulting from open data science tools helped us develop real working knowledge about the formulated problem and find the successful solution.

\section*{Acknowledgments}
Concerning applied computational methods and tools, we are grateful to Viktor Witkovsk\'y (Slovak Academy of Science, SK) for his recommendations and deep insights dealing with numerical inversion of characteristic functions and his MATLAB code
\texttt{CharFunTool}. We thank Aaron J. Hendrickson (US Department of Defense) for his insightful comments improving clarity of the article. Finally, we would also like to acknowledge the involvement and recommendations of Erik Bray (Université Paris-Sud, FR) and Luca de Feo (Université de Versailles Saint-Quentin-en-Yvelines, FR) in using SageMath, Jupyter, and GitHub in the frame of project OpenDreamKit (\url{https://opendreamkit.org/}).

\section*{Disclosure statement}

No potential conflict of interest was reported by the authors. 

\section*{Funding}

This work was supported by the Slovak Research and Development Agency under the contract no. APVV-17-0568 and the Internal Research Grant System of Faculty of Science, P. J. Šafárik University in Košice -- project vvgs-pf-2020-1423.

\section*{Notation}

\section*{ORCID}

Martina Han\v{c}ov\'a \orcidlink{} \url{https://orcid.org/0000-0001-8004-3972} \\
Andrej Gajdo\v{s} \orcidlink{}   \url{https://orcid.org/0000-0002-7004-6616} \\
Jozef Han\v{c} \orcidlink{} \url{https://orcid.org/0000-0003-1359-6117} \\

\bibliographystyle{tfnlm}
\bibliography{references,software}

\newpage

\appendix
\section{Used formulas}

\small
\textit{Generalized hypergeometric function} ${}_{p}F_{q}\left(\begin{smallmatrix}{\alpha_{1},\alpha_{2}, \ldots, \alpha_{p}} \\ {\beta_{1}, \beta_{2}, \ldots, \beta_{q}}\end{smallmatrix}\, ; z\right)$  
\\[6pt]
{\footnotesize
	\indent $\bullet$ as generalized hypergeometric series (DLMF \cite[eq. 16.2.1]{dlmf_nist_2021}); \\ \indent $\alpha_{1}, \alpha_{2}, \ldots, \alpha_{p} ; \beta_{1}, \beta_{2}, \ldots, \beta_{q} $ $\in \mathcal{R}$ or $\mathcal{C}$; Pochhammer's symbol $(\alpha)_n \equiv \alpha(\alpha+1)(\alpha+2)\ldots (\alpha+n-1) $}

\begin{equation} \label{form:genhypseries}
	\begin{gathered}
		{}_{p}F_{q}\left(\alpha_{1}, \alpha_{2}, \ldots, \alpha_{p} ; \beta_{1}, \beta_{2}, \ldots, \beta_{q} ; z\right) 
		= 
		\sum_{k=0}^{\infty} \tfrac{\left(\alpha_{1}\right)_{k}\left(\alpha_{2}\right)_{k} \cdots\left(\alpha_{p}\right)_{k}}{\left(\beta_{1}\right)_{k}\left(\beta_{2}\right)_{k} \ldots\left(\beta_{q}\right)_{k}} \frac{z^{k}}{k !} \\
		{}_{p}F_{q}\left(\alpha_{1}, \alpha_{2}, \ldots, \alpha_{p} ; \beta_{1}, \beta_{2}, \ldots, \beta_{q} ; z\right) \equiv 
		{}_{p}F_{q}\left(\begin{matrix}{\alpha_{1},\alpha_{2}, \ldots, \alpha_{p}} \\ {\beta_{1}, \beta_{2}, \ldots, \beta_{q}}\end{matrix}\, ; z\right) 
	\end{gathered}
\end{equation} 

\vertspace

\noindent \textit{Kummer's confluent hypergeometric function} $M(a,b,z)$ \\[3pt]
{\footnotesize 
	\indent $\bullet$ also known as $\Phi(a ; b ; z)$ or in the form of series \eqref{form:genhypseries} \\
	\indent \phantom{$\bullet$}  (Gradshteyn \& Ryzhik \cite[9.210]{gradshteyn_table_2007}, DLMF \cite[eq. 13.2.2]{dlmf_nist_2021}))}\\
\begin{equation} \label{form:Mabz}
	M(a,b, z)={}_{1} F_{1}\left(\begin{matrix} {a} \\ {b}\end{matrix}\, ; z\right)
\end{equation} \\[-6pt]

\noindent  \textit{Lower incomplete gamma function} $\gamma(a, x)$\\[3pt]
{\footnotesize
	\indent $\bullet$ as confluent hypergeometric function \eqref{form:Mabz} or in the form of series \eqref{form:genhypseries} \\
	\indent \phantom{$\bullet$} (DLMF, \cite[eq. 8.5.1]{dlmf_nist_2021}))
}
\begin{equation} \label{form:gamma}
	\begin{aligned}
		\gamma(a, x)&=a^{-1} x^{a} e^{-x} M(1,1+a, x)=a^{-1} x^{a} M(a, 1+a,-x) \\
		\gamma(a, z)&=a^{-1} z^{a} e^{-z} \, {}_{1} F_{1}\left(\begin{matrix}{1} \\ {a+1}\end{matrix} \, ; z\right)= a^{-1} z^{a}\,{}_{1} F_{1}\left(\begin{matrix}{a} \\ {a+1}\end{matrix}\, ;-z\right)
	\end{aligned}
\end{equation}

\vertspace

\noindent \textit{Error function}  $\operatorname{erf}(x)$ \\[3pt]
{\footnotesize 
	\indent $\bullet$ as $\gamma(a, x)$ or in the form of series \eqref{form:genhypseries} \\
	\indent \phantom{$\bullet$} (DLMF \cite[eqs. 7.11.1, 7.11.4]{dlmf_nist_2021})}
\begin{equation}\label{form:erf}
	\begin{gathered}
		\operatorname{erf}(x) \equiv \frac{2}{\sqrt{\pi}} \int_{0}^{x} e^{-t^{2}} d t = \frac{1}{\sqrt{\pi}}\,\gamma\left(\frac{1}{2}, x^{2}\right)\\
		\operatorname{erf}(z) = \dfrac{ 2 z e^{-z^{2}}}{\sqrt{\pi}}\, {}_{1}F_{1}\left(\begin{array}{c}{1} \\ {3/2}\end{array}; z^2\right) = \dfrac{ 2 z}{\sqrt{\pi}}\, {}_{1}F_{1}\left(\begin{array}{c}{1/2} \\ {3/2}\end{array}; -z^2\right)
	\end{gathered}
\end{equation}

\vertspace

\noindent \textit{Tricomi's confluent hypergeometric function} $U(a,b,z)$  \\[3pt]
{\footnotesize 
	\indent $\bullet$ also known as $\Psi(a ; b ; z)$ or in the form of series \eqref{form:genhypseries} \\
	\indent \phantom{$\bullet$}   (Gradshteyn \& Ryzhik \cite[9.210]{gradshteyn_table_2007}, Abramowitz \& Stegun \cite[13.1.10]{abramowitz_handbook_2014}) 
}
\begin{equation}\label{form:Uabz2F0}
	U(a,b,z)=z^{-a}{\,}_{2}F_{0}\left(a, a-b+1 ;-\dfrac{1}{z}\right)
\end{equation}

{\footnotesize 
	$\bullet$ integral representation of $U(a, b, z)$ \\
	\indent \phantom{$\bullet$}   (Gradshteyn \& Ryzhik \cite[9.211.4]{gradshteyn_table_2007}, DLMF \cite[eq. 13.4.4]{dlmf_nist_2021})}
\begin{flalign} \label{form:Uabzint}
&&	U(a, b, z)=\frac{1}{\Gamma(a)} \int_{0}^{\infty} e^{-z t} t^{a-1}(1+t)^{b-a-1} d t & \qquad[\Re a>0, \quad \Re z>0]
\end{flalign}

\newpage
\noindent \textit{Tricomi's confluent hypergeometric function} $U(a,b,z)$  \\[3pt]
{\footnotesize 
	\indent $\bullet$ derivative of $U(a, b, z )$ and values $U(a, b, 0 )$ \\
	\indent \phantom{$\bullet$}   (DLMF \cite[eqs. 13.2, 13.3.22]{dlmf_nist_2021})}
\begin{equation}
\begin{aligned} \label{form:Uab0}
	\hspace{2cm}U(a, b, 0)& =\frac{\Gamma(1-b)}{\Gamma(a-b+1)} & [\Re b<2, b\neq 0,1] \\
 U'(a,b,z) & \equiv \frac{d \,}{dz} U(a,b,z) = -aU(a+1,b+1,z) &
\end{aligned} 
\end{equation} \\

{\footnotesize 
	\indent $\bullet$ logarithmic derivative $\mathfrak{U}(a,b,z) = [\ln U(a,b,z)]'$ for $b\not\in \mathcal{Z}$ as a continued fraction\\
    \indent \phantom{$\bullet$} (Cuyt et al \cite[eq. 16.1.22]{cuyt_handbook_2008})}
\begin{equation}\mathfrak{U}(a,b,z)=-\frac{a}{z}+\frac{a(1+a-b) / z}{2 a-b+2+z} \dminus  
\bigK{m=1}{\infty}\left(\frac{(a+m)(b-a-m-1)}{b-2 a-2 m-2-z}\right) 
\end{equation}

\noindent  \textit{Whittaker's confluent hypergeometric functions $W_{\kappa, \mu}(z)$}  \\[3pt]
{\footnotesize 
	\indent $\bullet$ as Tricomi's function \eqref{form:Uabz2F0} or in the form series \eqref{form:genhypseries}\\
	\indent \phantom{$\bullet$}  (DLMF \cite[eq. 13.14.3]{dlmf_nist_2021})}
\begin{equation} \label{form:Wkmz}
	\begin{aligned}
		W_{\kappa, \mu}(z)=e^{-z / 2} z^{\mu+1 / 2} U\Big(\mu-\kappa+\tfrac{1}{2}, 2 \mu+1; z \Big) \\
		W_{\kappa, \mu}(z) = z^{\kappa} e^{-z/2}\,{}_{2} F_{0}\Big(-\kappa-\mu+\tfrac{1}{2},-\kappa+\mu+\tfrac{1}{2} ;-\tfrac{1}{z}\Big)
	\end{aligned}
\end{equation}

{\footnotesize 
	$\bullet$ integral identities including  $W_{\kappa, \mu}(z)$ \\
	\indent \phantom{$\bullet$}   (Gradshteyn \& Ryzhik \cite[3.383.4]{gradshteyn_table_2007}}
\begin{equation}
\begin{aligned} \label{form: WkzGDDconv}
		\int_{u}^{\infty} x^{\nu-1}(x-u)^{\mu-1} e^{-\beta x} d x=\beta^{-\frac{\mu+\nu}{2}} u^{\frac{\mu+v-2}{2}} \Gamma(\mu) \exp \left(-\frac{\beta u}{2}\right) W_{\frac{\nu-\mu}{2}, \frac{1-\mu-\nu}{2}}(\beta u) \\
	 [\Re \mu>0, \quad \Re \beta u>0]
\end{aligned}
\end{equation}
{\footnotesize 
\indent \phantom{$\bullet$}   (Gradshteyn \& Ryzhik \cite[3.384.9]{gradshteyn_table_2007})}
\begin{equation} \label{form: Wkznuminv}
	\begin{multlined}
		\int_{-\infty}^{\infty}(\beta+i x)^{-2 \mu}(\gamma-i x)^{-2 \nu} e^{-i p x} d x  =\\
		=2 \pi(\beta+\gamma)^{-\mu-\nu} \frac{p^{\mu+\nu-1}}{\Gamma(2 \nu)} \exp \left(\frac{\beta-\gamma}{2} p\right) W_{\nu-\mu, \frac{1}{2}-\nu-\mu}(\beta p+\gamma p) \,\quad\qquad[p>0] \\
		=2 \pi(\beta+\gamma)^{-\mu-\nu} \frac{(-p)^{\mu+\nu-1}}{\Gamma(2 \mu)} \exp \left(\frac{\beta-\gamma}{2} p\right) W_{\mu-\nu, \frac{1}{2}-\nu-\mu}(-\beta p-\gamma p) \quad[p<0] \\
		{\left[\Re \beta>0, \quad \Re \gamma>0, \quad \Re(\mu+\nu)>\tfrac{1}{2}\right]}
	\end{multlined}
\end{equation} \\
\noindent \textit{Gauss hypergeometric function} $F(\alpha, \beta;\gamma;z)$  \\[3pt]
{\footnotesize 
	\indent $\bullet$ as series \eqref{form:genhypseries} (Gradshtyen \& Ryzhik \cite[9.100, 9.14.2]{gradshteyn_table_2007}, DLMF \cite[eq. 15.2.1]{dlmf_nist_2021})}
\begin{equation}\label{form:gausshypgeo}
	F(\alpha, \beta;\gamma;z) = {}_2F_1\left(\begin{matrix}
		\alpha, \beta \\
		\gamma
	\end{matrix}\,; z\right) 
\end{equation} 

{\footnotesize 
	\indent $\bullet$ integral identity between $\gamma(a, x)$ and $F(\alpha, \beta;\gamma;z)$ \\
	\indent \phantom{$\bullet$}  (Gradshteyn \& Ryzhik \cite[6.455.2]{gradshteyn_table_2007})
}
\begin{equation} \label{form:gammagauss}
	\begin{aligned}
		\int_{0}^{\infty} x^{\mu-1} e^{-\beta x} \gamma(\nu, \alpha x) \,d x  = \frac{\alpha^{\nu} \Gamma(\mu+\nu)}{\nu(\alpha+\beta)^{\mu+\nu}} \,	F\left(
			1, \mu+\nu ;	\nu+1; \tfrac{\alpha}{\alpha+\beta}\right) \\
		[\Re(\alpha+\beta)>0, \quad \Re\beta>0, \quad \Re(\mu+\nu)>0]
	\end{aligned}
\end{equation}

\newpage
\section{$\GDD$ in times series kriging}

\begin{figure}[H]
	\captionsetup{font=small}
	\centering
	\includegraphics[width=0.8\textwidth]{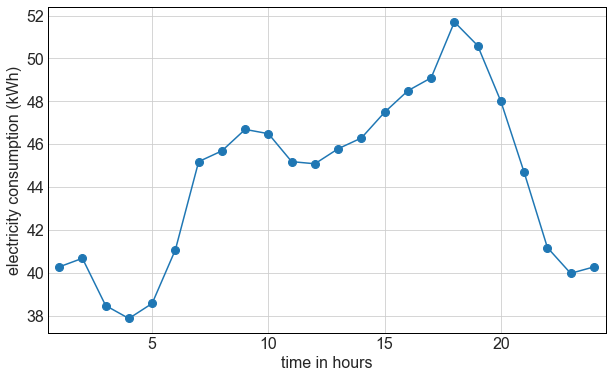}
	\caption{Time series data of electricity consumption during 24 hours in a department store.}
	\label{fig:timeseries}
\end{figure}

As a real data example,  which becomes the basis for our numerical study, we use a microeconomic time series dataset 
(fig. \ref{fig:timeseries}), analyzed and modeled in \cite{gajdos_kriging_2017,hancova_estimating_2020}. The time series data set is available in a Jupyter notebook at our GitHub repository \cite{gajdos_fdslrm:_2019}\footnote{Since this time series data set, first published in \cite{stulajter_estimation_2004}, contains too few values regarding the number of model parameters, we consider it as a testing touchstone example of how models in the time series kriging look like. On the other hand, its simplicity allows us to carry out effective simulation and numerical studies and compare various results over years.}. 

The Gaussian time-series model $X(.)$ suitably fitting the electricity data in the frame of kriging leads to the following linear mixed model
\begin{equation}\label{eq:fdslrm}
\Xv = \F\bv +\V\Yv+\Zv, \quad  \Cov{\Xv}   = \s_0^2\II_{24}+\V\D\V' \equiv \Sg.
\end{equation} 
\noindent where\\

\indent $\Xv = (X(1),\ldots,X(24))'  \sim \NnX, \,\, \bv=(\beta_1,\,\beta_2,\,\beta_3)'\in\mathcal{R}^3,$ \\[6pt]
\indent $\Zv = (Z(1),\ldots,Z(24))'\sim \NX_{24}(\Ov{24}, \s_0^2\II_{24}), \,\,\Yv = (Y_1, Y_2, Y_3, Y_4)' \sim \mathcal{N}_4(\boldsymbol{0}_4, \D),$ \\[6pt]
\indent $ \Cov{\Yv,\Zv} = \Om{4}{24}, \,\, \D = \Diag{(\s_1^2, \s_2^2, \s_3^2, \s_4^2)'},  \,\, \nuv = (\sigma_0^2, \sigma_1^2, \sigma_2^2, \sigma_3^2, \sigma_4^2)' \in \mathcal{R}_{+}^5.$ \\

Since econometric datasets almost always show some periodic patterns as they are influenced by seasons or regularly repeating events, the model fitting procedure in kriging is based on spectral analysis \cite{gajdos_kriging_2017}. Design matrices $\F, \V$ are therefore given by the three most significant Fourier frequencies  $\left(\omega_{1}, \omega_{2}, \omega_{3}\right)'=2 \pi(1 / 24,2 / 24,3 / 24)'$ as
$$
\begin{gathered}
\small
\F\equiv  \big(\fv_1, \fv_2,\fv_3\big) = \\
\footnotesize
\left(\begin{array}{lll}
1 & \cos\left(\omega_{1}\right) & \sin\left(\omega_{1}\right) \\
1 & \cos\left(2 \, \omega_{1}\right) & \sin\left(2 \, \omega_{1}\right) \\
\vdots & \vdots & \vdots \\
1 & \cos\left(24 \, \omega_{1}\right) & \sin\left(24 \, \omega_{1}\right)
\end{array}\right) 
\end{gathered} 
\quad
\quad
\begin{gathered}
\small
\V  \equiv  \big(\vv_1, \vv_2,\vv_3, \vv_4\big) =  \\
\footnotesize
\left(\begin{array}{llll}
\cos\left(\omega_{2}\right) & \sin\left(\omega_{2}\right) & \cos\left(\omega_{3}\right) & \sin\left(\omega_{3}\right) \\
\cos\left(2 \, \omega_{2}\right) & \sin\left(2 \, \omega_{2}\right) & \cos\left(2 \, \omega_{3}\right) & \sin\left(2 \, \omega_{3}\right) \\
\vdots & \vdots & \vdots & \vdots \\
\cos\left(24 \, \omega_{2}\right) & \sin\left(24 \, \omega_{2}\right) & \cos\left(24 \, \omega_{3}\right) & \sin\left(24 \, \omega_{3}\right)
\end{array}\right)
\end{gathered}
$$
satisfying the following condition of orthogonality
$$
\small
\begin{gathered}
\F'\V = \Om{3}{4}, \F'\F = \Diag{\big(\norm{\fv_1}^2\!\!,\norm{\fv_1}^2\!\!,\norm{\fv_3}^2\big)' } = 
\Diag{\big(24,12,12\big)'}, \\[-3pt] 
\V'\V = \Diag{\big(\norm{\vv_1}^2\!\!,\norm{\vv_2}^2\!\!,\norm{\vv_1}^3\!\!, \norm{\vv_4}^2 \big)'} = 
\Diag{\big(12,12,12,12\big)'}.
\end{gathered}
$$

There are several methods \cite{hancova_estimating_2020}, based on least squares or maximum likelihood,
how to estimate unknown time series model parameters: $k$ regression parameters $\bv$ ($k=3$) and $l$ non-negative variance parameters $\nuv$ ($l=4$). 
For example the non-interative method of moments gives us unbiased quadratic estimators $\mmes{j}; j = 1,..,l$ (called \textit{MM estimators}) as 
\begin{equation}
\begin{gathered} \label{eq:MMest}
\mmes{j} = \nes{j} - \tfrac{1}{\norm{\vv_j}^2}\nes{0}, \quad \nes{0} = \tfrac{1}{n-k-l}{\Xv' \M \Xv}, \quad \nes{j} = \tfrac{1}{\norm{\vv_j}^4} \Xv' \vv_j\vv_j'\Xv,
\end{gathered}
\end{equation}
where $\M$ is the orthogonal projection matrix to column space $(\F\,\, \V)$. The electricity data produce the following estimations of $\nes{0}$, $\nes{j}$:
\begin{equation}\label{eq:MMvalues}
\tilde\nuv = (\nes{0}, \nes{1}, \nes{2}, \nes{3}, \nes{4})'  =  (1.09, 2.97, 1.76, 0.37, 1.86)'.
\end{equation}
As for distribution of MM estimators, standard distribution theorems for quadratic forms (see e.g. \cite[ch. 10.5]{searle_matrix_2017}) together with basic properties of gamma distribution \cite{mittelhammer_mathematical_2013} and algebraic properties of mutually orthogonal matrices $\M, \V$ \cite{hancova_natural_2008} give us
$$\small\Xv' \M \Xv \sim \Gm\left(\dfrac{n-k-l}{2},\dfrac{1}{2\s^2_0}\right), \quad \Xv' \vv_j \vv_j \Xv \sim \Gm\left(\dfrac{1}{2},\dfrac{\norm{\vv_j}^2}{2(\s^2_0+\norm{\vv_j}^2\s^2_j)}\right).$$
These quadratic forms are also independent, so   we get the $\GDD$ distribution of MM estimators for our model
\begin{equation}
\begin{gathered} \label{eq:GDDMMest}
\mmes{j} \sim \GDD(\a_1,\b_1,\a_2,\b_2) \, \text{ for } \, j = 1, 2, 3, 4 \\
\small
\a_1=\frac{1}{2},\, \b_1= \frac{\norm{\vv_j}^2}{2(\s^2_0+\s^2_j\norm{\vv_j}^2)},\, 
\a_2=\frac{n-k-l}{2},\, \b_2=\frac{(n-k-l)\norm{\vv_j}^2}{2\s^2_0},
\end{gathered}
\end{equation}
where substitution $n = 24, k=3, l=4$ provides $\GDD$ parameters in the form
\begin{equation*}
\small
\a_1=\frac{1}{2},\, \b_1= \frac{6}{\s^2_0+12\s^2_j},\, 
\a_2=\frac{17}{2},\, \b_2=\frac{102}{\s^2_0}.
\end{equation*}
Finally, using estimations $\nuv$ from \eqref{eq:MMvalues}, we get for $\GDD$ parameters
\begin{equation*}
\begin{gathered}
\a_1 = 1/2, \,  \a_2 = 17/2, \, \a = \a_1+\a_ 2= 9,\\[6pt]
\b_1 \in \{0.16, 0.27, 1.08, 0.26\}, \b_2 = 93.32, \\
\b = \b_1+\b_2  \in \{93.48, 93.59, 94.40, 93.58\}. 
\end{gathered}
\end{equation*}

\noindent Based on the previous result, for the sake of simplicity, we take the following set of typical $\GDD$ parameters in our numerical study 
\begin{equation} \label{eq:GDDtypical}
	\begin{gathered}
		\a_1 = 0.5, \,  \a_2 = 8.5, \, \a = \a_1+\a_ 2= 9.0, \\[3pt]
		\b_1 = 1.0, \b_2 = 93.0, \b = \b_1+\b_2  = 94.0. 
	\end{gathered}
\end{equation}

\newpage
\section{Open digital tools in the numerical study}

\begin{tcolorbox}[colback=white]
	Box C1: Chosen open digital tools for our numerical study \\[-18pt]
	
	\noindent\hrulefill\\[-6pt]
	
	{\small
		\MakeUppercase{Open digital tools} \\[6pt]
		\indent \textbf{SageMath} \cite[v.~\!9.2,]{sage}, \textbf{Python} \cite[v.~\!3.7.7,][]{python3}, \textbf{R} \cite[v.~\!4.0.5,][]{Rcore}, \textbf{Octave} \cite[v.~\!6.2.0,][]{octave}
		\begin{itemize}
			\setlength{\itemindent}{-12pt}
			\setlength\itemsep{3pt}
			\item \textit{numerical and scientific computing libraries }\\[-6pt]
			\begin{itemize}
				\setlength{\itemindent}{-30pt}
				\setlength\itemsep{3pt}
				\item \textbf{NumPy} \cite[v.~\!1.19.1,][]{numpy2020} numerical Python library for fast vector computations
				\item \textbf{SciPy} \cite[v.~\!1.5.2,][]{scipy} fundamental Python library for scientific computing
				\item \textbf{GSL }\cite[v.~\!2.6,][]{galassi_gnu_2009}: C, C++ numerical library for scientific computing
				\item \textbf{PARI/GP} \cite[v.~\!2.11.4,][]{pari}: CAS and C arbitrary-precision library
				\item \textbf{mpmath} \cite[v.~\!1.1.0,][]{mpmath}: arbitrary-precision Python library 
				\item \textbf{Arb} \cite[v.~\!2.16.0,][]{johansson_arb_2014}: C library for arbitrary-precision interval arithmetic using the midpoint-radius representation ("ball arithmetic")
			\end{itemize}
			\item \textit{high performance Python compilers }\\[-6pt]
			\begin{itemize}
				\setlength{\itemindent}{-30pt}
				\setlength\itemsep{3pt}
				\item \textbf{Cython} \cite[v.~\!0.29.21,][]{cython}: optimizing static compiler translating Python into C code
				\item \textbf{Numba} \cite[v.~\!0.53.1,][]{numba2015}: JIT compiler translating Python into fast machine code
			\end{itemize}
		\end{itemize}
	
}\end{tcolorbox} 

\vspace{-12pt}

\subsubsection*{Open data science tools}

R and Python with NumPy, SciPy are currently the main open data science tools and scientific computing standards in many fields. Moreover today Python allows to use
high-performance compilers Numba and Cython, which can accelerate Python code performance to the speeds of Fortran, C or C++.
In applying these compilers, we were inspired by \cite{boulle_high-performance_2019}.

One of the highly appreciated Numba compiler features is its very simple way of code optimization --- adding only one extra line to a pure Python code. This line applies the so-called Numba decorator, e.g. we use \texttt{@njit(fastmath=True)}). Then Numba generates a fast machine code. Cython is a powerful combination of Python language that additionally allows calling C functions and declaring C types on Python objects (e.g. using complex variable x means to add extra line \texttt{cdef double complex x}). It means that we combine easily readable Python code with adding C types of variables, also in Python syntax. The Cython result is an efficient, fast C code. 

\vspace{-6pt}
\subsubsection*{Open Python-based SageMath}
Since our numerical study requires analytic and numerical computations together with programming allowing automated and easy reproducible benchmarking, we also use free open Python-based mathematical software SageMath \cite{stein_sage_2019,zimmermann_computational_2018}, shortly Sage, running in Jupyter environment. Sage is a \textit{defacto} sophisticated Python library that allows running all mentioned Python tools. Today Sage, which is also a computer algebra system (CAS), is considered as a viable alternative to commercial scientific computing tools like Mathematica \cite{wolfram}, Maple \cite{maple}, or MATLAB \cite{matlab}. The uniqueness of SageMath lies especially in the possibility to run, control, and communicate with many other powerful third-party computing tools through a common, very accessible Python-based language, all in one environment (Jupyter notebook).

\vspace{-6pt}
\subsubsection*{Arbitrary-precision libraries}
Sage is built on top of powerful open third-party computing tools, e.g. in numerical integration Sage implemented C library GSL or for arbitrary-precision aritmetics library mpmath. Sage also allows running of very powerful and highly precise C libraries PARI/GP and Arb. Libraries mpmath, PARI/GP and Arb can be used for generating and simultaneous cross-check of high-precision values of $f(x)$ using $U(a,b,z)$ \eqref{eq:GDD_densU}, which can be subsequently used for computing real numerical errors in results for all used computational tools. 

\end{document}

%% file: OwnSymbolsCommands.tex
\usepackage{graphicx}
\usepackage{amsmath}
\usepackage{amsfonts}
\usepackage{amssymb}
\usepackage{epstopdf}
\usepackage[utf8]{inputenc}
\usepackage[english]{babel}
\usepackage{upgreek}
\usepackage{dsfont}
\usepackage{mathrsfs} 
\usepackage{orcidlink}
\usepackage{ctable}
\usepackage{float}	
\usepackage{mathbbol}
\usepackage[justification=centering]{caption}
\usepackage{enumitem}
\usepackage{array,multirow}
\usepackage{caption}
\usepackage{hyperref}
\hypersetup{
	colorlinks=true,
	linkcolor=blue,
	filecolor=magenta,      
	urlcolor=blue,
	citecolor=red,
}
\usepackage{tcolorbox}

\newcommand{\Section}[1]{\section{#1} \setcounter{equation}{0}}

\newcommand{\MMA}{\textit{MMA}}
\newcommand{\MTB}{\textit{MTB}}

\newcommand{\mathbi}[1]{\boldsymbol{#1}} 

\renewcommand{\a}{\upalpha}
\renewcommand{\b}{\upbeta}

\newcommand{\BigK}{\mathop{\mbox{\rm \huge K}}}
\newcommand{\bigK}[2]{\mbox{\raisebox{-5pt}{$\displaystyle\BigK_{#1}^{#2}$}}}

\newcommand{\dminus}{\mathrel{\raisebox{-7pt}{$-$}}}

\newcommand{\cf}[1]{\varphi_X(#1)}
\newcommand{\cfi}[1]{\varphi(#1)}

\newcommand{\vertspace}{\vspace{6pt}}

\newcommand{\GDD}{\mathcal{GDD}}
\newcommand{\Gm}{\mathcal{G}}

\newcommand{\mi}[1]{\mathit{#1}}
\renewcommand{\x}{\times}
\newcommand{\xm}{\times\,\!}
\newcommand{\epsr}{\varepsilon_{\text{rel}}}
\newcommand{\bv}{\mathbi{\beta}}

\newcommand{\Yv}{\mathbi{Y}}
\newcommand{\Xv}{\mathbi{X}}

\newcommand{\Zv}{\mathbi{Z}}

\newcommand{\Ov}[1]{\boldsymbol{0}_{#1}}
\newcommand{\Om}[2]{\boldsymbol{0}_{#1 \times #2}}

\newcommand{\fv}{\mathbi{f}}
\newcommand{\vv}{\mathbi{v}}

\newcommand{\E}[1]{E\left\{#1\right\}}

\newcommand{\nuv}{\boldsymbol{\upnu}}

\newcommand{\mode}{\mathcal{M}}
\newcommand{\s}{\sigma}                         
\newcommand{\Sg}{\mathbf{\Sigma}}

\newcommand{\Cov}[1]{{Cov}\{#1\}}

\newcommand{\M}{\mathbf{M}}
\newcommand{\V}{\mathbf{V}}
\newcommand{\F}{\mathbf{F}}

\newcommand{\pitem}{$(\arabic*)$}

\newcommand{\mmes}[1]{\hat{\s}^2_{#1}}

\newcommand{\nes}[1]{\tilde{\s}_{#1}^2}

\newcommand{\NnX}{\mathcal{N}_{24}(\F\bv, \Sg)}
\newcommand{\NX}{\mathcal{N}}

\newcommand{\Rv}[1]{\mathcal{R}^{#1}}
\newcommand{\C}{\mathcal{C}}
\newcommand{\R}{\mathcal{R}}

\newcommand{\bigO}{\mathcal{O}}

\newcommand{\D}{\mathbf{D}}

\newcommand{\II}{\mathbf{I}}

\newcommand{\norm}[1]{\left\Vert#1\right\Vert}

\newcommand{\Diag}[1]{\mathrm{diag}\left\{#1\right\}}

\newcommand{\Sum}[2]{\sum\limits_{#1=1}^{#2}}

